\documentclass[aps,a4paper,superscriptaddress,preprint,showpacs,showkeys,12pt,eps]{revtex4}

\usepackage{graphicx}
\usepackage{subfigure}
\usepackage{latexsym}   

\begin{document}

\title{Phase transitions and autocorrelation times in two-dimensional Ising model with
dipole interactions}

\author{\firstname{Leandro} G. \surname{Rizzi}}
\email{lerizzi@pg.ffclrp.usp.br} 
\author{\firstname{Nelson} A. \surname{Alves}}
\email{alves@ffclrp.usp.br}

\affiliation{Departamento de F\'{\i}sica e Matem\'{a}tica, FFCLRP, \\
             Universidade de S\~ao Paulo,
             Avenida Bandeirantes, 3900 \\ 
             14040-901, Ribeir\~ao Preto, SP, Brazil.}
\date{\today}

\begin{abstract}
   The two-dimensional Ising model with nearest-neighbor ferromagnetic and long-range dipolar interactions exhibits a rich phase diagram. 
   The presence of the dipolar interaction changes the ferromagnetic ground state expected for the pure Ising model to a series of striped phases as a function of the interaction strengths. 
   Monte Carlo simulations and histogram reweighting techniques
applied to multiple histograms are performed to identify the critical temperatures for the phase transitions taking place for stripes of width $h=2$ on square lattices.
   In particular, we aim to study the intermediate nematic phase, which is observed for large lattice sizes only.
   For these lattice sizes, we calculate the critical temperatures for the striped-nematic and nematic-tetragonal transitions, critical exponents, and the bulk free-energy barrier associated with the coexisting 
phases.
   We also evaluate the long-term correlations in our time series near the finite-size critical points by studying the integrated autocorrelation time $\tau$ as a function of the lattice size.
  This allows us to infer how severe the critical slowing down for this system with long-range interaction and nearby thermodynamic phase transitions is.

\end{abstract}


\maketitle


\maketitle

\section{Introduction}

  Simple two-dimensional (2D) Ising-like models may present a complex behavior with rich phase diagrams. 
  A complex behavior may appear even in face of simple competing interactions.
  In general, a complex behavior results from the frustration phenomenon caused by these competing interactions
at different length scales. 
  
  The 2D Ising model with nearest-neighbor ferromagnetic exchange and dipolar interaction has been the focus of considerable interest because of its magnetic properties \cite{debell_72_2000}.
  The Hamiltonian model is defined by the Ising lattice model with nearest-neighbor exchange interaction $J> 0$ and dipolar contribution of strength $g>0$,
which is rewritten as 
\begin{equation}
          {\cal H} = -\delta \sum_{<i,j>} \sigma_{i} \sigma_{j} +  
                      \sum_{i < j} \frac{\sigma_{i} \sigma_{j}}{r_{ij}^{3}} \, ,     \label{eq:hamiltonian}
\end{equation}
where $\delta = J/g$.
  Here, we have adopted the convention \cite{cannas_68_2003} of summing over all distinct pairs of lattice spins at distances $r_{ij}$ to define the constant $g$. 
  The distances $r_{ij}$ are measured in units of lattice.

  The competition between the exchange and dipolar interactions
results in a phase diagram that is rather sensitive to the $ \delta = J/g$ ratio \cite{cannas_75_2007,Rastelli_PRB76_2007}.
  For $\delta < 0.4403$, the model presents antiferromagnetic ground-state characterized by
stable checkerboard like configurations \cite{Rastelli_PRB76_2007}.
  For larger $\delta$ values, the ground state changes to striped phases.
  These phases are characterized by magnetic domains displayed in stripes of alternating spins
perpendicular to the lattice plane.
  Each stripe corresponds to a pattern of upward or downward spins, whose 
width $h$ increases with $\delta$ \cite{macisaac_51_1995,giuliani_76_2007}.  
  As the temperature increases, for fixed couplings, an order-disorder transition takes place until the system reaches its usual paramagnetic phase.
  
  This theoretical aspect has been observed in experiments with ultrathin 
(i.e., a few atomic layers) metal films on metal substrates \cite{debell_72_2000,portmann_422_2003}
below certain temperatures as a consequence of a reorientation transition of their spins.   
  Apart from short-range interactions, long-range dipolar interactions 
are also present between the magnetic moments in these systems.
  As a matter of fact, even for the simple ferromagnetic Ising model, one should always include 
 a long-range $1/r^3$ repulsive interaction between the magnetic moments
in addition to the short-range exchange interaction term.
  Theoretical studies of magnetic systems have avoided the inclusion of the relatively small
dipolar interaction compared with the exchange interaction. 
  However, as we have mentioned the dipolar interaction plays an important role 
for quasi 2D-magnetic systems in characterizing metal films \cite{debell_72_2000}.
  Furthermore, from a theoretical viewpoint, a long-range repulsive interaction $1/r^s$
for $d$-dimensional systems, with $d < s \leq d+1$, in addition to the usual nearest-neighbor ferromagnetic interaction, leads to the absence of ferromagnetism for all
temperatures \cite{giuliani_76_2007,biskup_274_2007}.
  
 Efforts have been made to determine the phase diagram as a function of the coupling $\delta$
from mean field approximation and numerical simulations
\cite{cannas_68_2003,cannas_75_2007,macisaac_51_1995,cannas_168_2002}. 
  Analytical and Monte Carlo (MC) calculations have shown the following picture:
  For a fixed coupling $\delta > 0.4403$, the phase diagram initially exhibits the alternating striped spin configurations as described above, characterizing the so called smectic striped phase.
  As the temperature increases, the initial picture describes a transition into the
tetragonal phase \cite{ibooth_75_1995,cannas_69_2004,casartelli_37_2004,arlett_54_1996}
characterized by states with orientationally disordered stripes but still preserving some of this structural form, which tends to the completely disordered paramagnetic state. 
  Recent numerical results related to this striped-tetragonal phase transition
indicate a continuous transition for $h=1$ \cite{rastelli_73_2006}, a clear first-order transition for $h=2$ \cite{cannas_69_2004,rastelli_73_2006}, a likely weaker first-order transition for $h=3$ \cite{rastelli_73_2006}, and a continuous transition for $h=4$ \cite{rastelli_73_2006} and $h=8$ \cite{casartelli_37_2004}. 
   On the other hand, a theoretical approach developed by Abanov {\it et al.} \cite{abanov_54_1995},
based on a continuum limit for the Hamiltonian, has predicted a new domain phase in the phase diagram.
   In addition to the striped and tetragonal phases, there would be an intermediate Ising nematic phase,
whose existence would depend on the parameters of the model.
      
   Different methods have been used to study the criticality of this model.
   As summarized above, the order of the phase transition between the low temperature smectic striped 
and the high temperature tetragonal phases is still under debate and seems to depend on the 
coupling $\delta$.
   Simulations have been hampered because large lattice size simulations are very CPU timeconsuming due to the dipolar term.
   In spite of these limitations, recent literature works have presented very satisfactory results,
including numerical evidence about the intermediate Ising nematic phase \cite{cannas_75_2007,cannas_73_2006}.
   The conclusion seems to be that the nematic phase can be observed in a narrow range of temperatures and is $\delta$-dependent.
   This amounts to observing not only a single peak in the specific heat,
initially corresponding to the expected striped-tetragonal phase transition, but a richer phase diagram
comprising the striped-nematic and the Ising nematic-tetragonal transitions.
   This new scenario has been identified, in particular, for $\delta=2$, corresponding to stripes of width $h=2$
in the ground state. 
    
   In this paper, we perform extensive MC simulations to study the phase transitions involved in this model for $\delta=2$. 
   We present convincing numerical evidence for an Ising nematic phase for this coupling
and discuss the calculation of the critical exponents in face of the constraint related to lattice sizes necessary for 
observation of the true phenomenology of this model. 
   Our study also allows us to figure out how reliable the simulations performed with a local 
update algorithm are.
   Using histogram reweighting techniques \cite{ferrenberg_61_1988,alves_376_1990} and the patching procedure \cite{alves_376_1990} to combine data obtained from simulations at various temperatures, we evaluate the specific heat and the order parameter susceptibility over a wide range of temperatures, and the free-energy barrier associated with the coexisting phases.
   This analysis procedure enables us to explore the existence of distinct maxima as a continuous function of the temperature and thus, to estimate the finite-size critical temperatures with high precision.
   Our results show that simulations need to be performed for lattice sizes $L$ at least
as large as $L=48$ so that the Ising nematic phase can be clearly observed.
   Consequently, the conclusions presented in the literature for smaller lattice sizes can be misleading.
   Moreover, to ensure the correct determination of the physical quantities
and their error bars, one needs to have a reasonable number of independent measurements.
   Thus, to access how the long-range dipolar interaction may affect physical estimates, we also evaluate the integrated autocorrelation time in energy time series \cite{sokal_89}.

   In Sec. 2, we briefly review the histogram reweighting and patching techniques 
which we use to estimate average energies, orientational order parameter and their
response functions, specific heat and susceptibility, as a function of temperature. 
   Formation of the nematic phase is clearly exhibited as we increase the lattice size.
   In all cases, we estimate the specific heat maxima and finite-size critical temperatures by including Jackknife error estimates in the analysis \cite{Efron}. 
   Our histograms for the energy distributions present double-peak structure leading to the existence of domain walls between the states of each phase.
   Thus, we also evaluate the free-energy barriers at both phase transitions.
   The integrated autocorrelation time analysis is presented in Sec. 3, which is followed by
a brief outlook and our conclusions in Sec. 4.

\section{Monte Carlo simulations}

   We consider the two-dimensional system described by the Hamiltonian in Eq. (\ref{eq:hamiltonian}), where the spin variables $\sigma=\pm 1$ are assumed to be aligned perpendicular to the square lattice with $N=L^2$ sites.
   We have used the standard Metropolis algorithm to generate configurations for lattice
sizes $L=16, 32, 48, 56, 64$ and $L=72$, with $\delta=2$. 
   We carried out our simulations with periodic boundary conditions to minimize border effects due
to the dipolar term in the Hamiltonian. 
  Thus, all distances $r_{ij}$  must include sites in the infinitely replicated simulation box in both directions. 

  The boundary conditions add an infinite sum over all images of the simulation box to the dipole term of the Hamiltonian.  
  We use Ewald summation technique to compute this infinite sum. 
  This technique splits the infinite sum over all images of the system into two quickly converging sums: the direct sum, which is evaluated in the real space, the reciprocal sum, carried out in the reciprocal space, and a self-interaction correction term \cite{CompPhysComm.95.1996, JChemPhys.106.1997}. 
  We have set the Ewald parameter $\alpha$ to 3.5 in all simulations.
  This parameter determines the rate of convergence between the two sums.

  Our MC Markov chains are initialized with random starts. 
  For thermalization, $10^6$ sweeps were discarded for $L \leq 48$, and 
$5 \times 10^5$ sweeps for $L = 56, 64$ and 72.
  Our checks have shown that after $\sim10^{5}$ sweeps the system does not present considerable changes in the energy. A detailed analysis about the time evaluation can be found in Ref. \cite{cannas_78_2008}.
  For each temperature, measurements rely on $3.4 \times 10^7$ sweeps for $L \leq 64$.
  Our largest lattice size $L =72$ relies only on $2.7 \times 10^7$ sweeps because even a 10 times larger
simulation would not present a reasonable number of independent measurements in face of our estimates from the
integrated autocorrelation time calculations for smaller lattice sizes depicted in Fig. 8.
  However, the patching procedure attenuates this drawback because we simultaneously match
simulations at different temperatures $T_0$ to obtain the final physical estimates.
 
\begin{table}[t]
\begin{center}
\begin{tabular}{@{}llllll}
\hline \hline
~$L$ &  ~~~~~~~~~~~~~~~~~~~~$T_0$  & $C_v|_{\rm max}$  & ~~~$T_{\rm max}^{(1)}$ & 
                                               $C_v|_{\rm max}$  & ~~~$T_{\rm max}^{(2)}$ \\
\hline
 16 & 0.791, 0.830, 0.850, 0.870               &         &          & 3.095(2) & 0.8323(2) \\
 32 & 0.780, 0.791, 0.812, 0.825, 0.850        &         &          & 3.92(2)  & 0.7905(3) \\
 48 & 0.770, 0.780, 0.791, 0.812, 0.870        & 3.76(3) & 0.7785(7)& 4.21(1) & 0.8132(2)  \\
 56 & 0.760, 0.770, 0.780, 0.790, 0.808, 0.860 & 3.72(4) & 0.7727(6)& 5.44(2) & 0.8084(2)  \\
 64 & 0.767, 0.773, 0.780, 0.800, 0.808        & 3.76(5) & 0.7726(7)& 6.05(3) & 0.8045(2)  \\
 72 & 0.760, 0.770, 0.780, 0.790, 0.807, 0.830 & 4.05(35)& 0.767(2) & 5.63(14)& 0.800(1)  \\
\hline \hline
\end{tabular}
\caption{ Data produced at temperatures $T_0$ and used for patching and reweighting. 
$T_{\rm max}$: finite-size critical temperatures for the phase transitions observed for system sizes $L$ defined by the maximum of specific heat $C_v|_{\rm max}$ as shown in figures 1(a) and 1(b).}
\end{center}
\end{table}

  To identify the finite-size critical temperatures, we evaluate the specific heat
\begin{equation}
    C_v(T) \ =\ \frac{1}{T^2 N} (\langle E^2 \rangle - \langle E\rangle^2)         \label{cv}
\end{equation}
over a continuous range of temperatures by reweighting MC data
obtained from simulations at $T_0$. 
  The list of temperatures and our final estimates for the specific heat maxima $C_v|_{\rm max}$ and their 
  finite-size critical temperatures $T_{\rm max}$ is displayed in table 1.
  All error bar estimates rely on 40 Jackknife bins for $L \leq 64$ and 20 bins for $L=72$.
   
  With the time series produced at temperatures $T_0^{(i)}$, ($i=1, \cdots, P$),  
histogram reweighting techniques allow us to evaluate physical quantities in neighborhoods
of the simulated temperatures,
\begin{equation}
  \overline{f(T)}  = 
  \frac{1}{Z(T)} \sum_{n=1}^{nmeas} \,f_n\, {\rm exp}\left[\Delta(\frac{1}{T}) E_n\right] \, ,
\end{equation}
where $nmeas$ is the total number of measurements in both energy $E$ and
physical quantity $f$ time series; $\Delta(1/T) = 1/T -1/T_0$, and $Z(T)$ is the partition function,
\begin{equation}
 Z(T) = \sum_{n=1}^{\rm nmeas} {\rm exp}\left[\Delta(\frac{1}{T}) E_n\right] \, .
\end{equation}


\begin{figure}[!t]
\begin{center}
\begin{minipage}[h]{0.50\textwidth}
\subfigure{\includegraphics[width=0.95\textwidth]{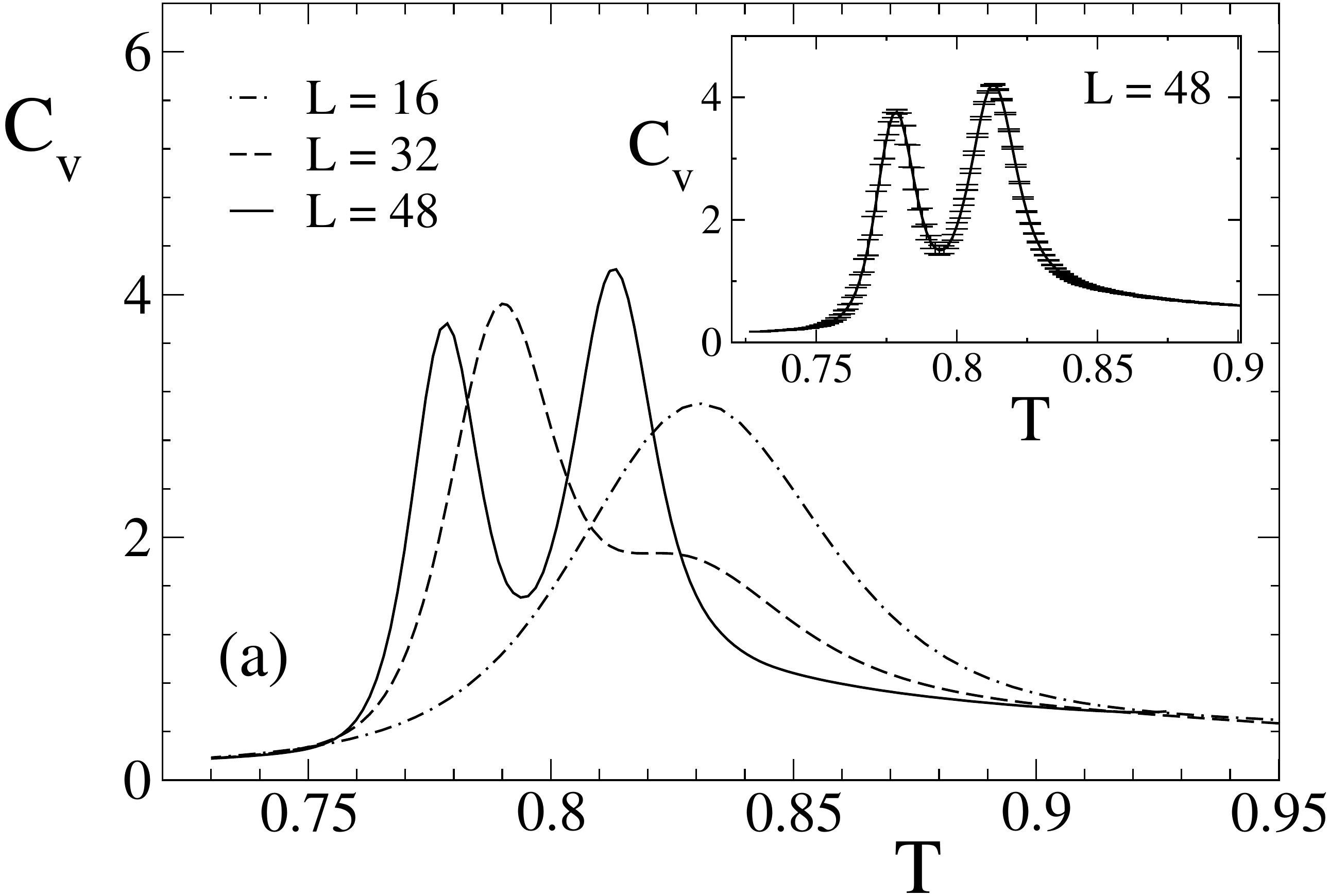}}
\end{minipage}%
\begin{minipage}[h]{0.50\textwidth}
\subfigure{\includegraphics[width=0.95\textwidth]{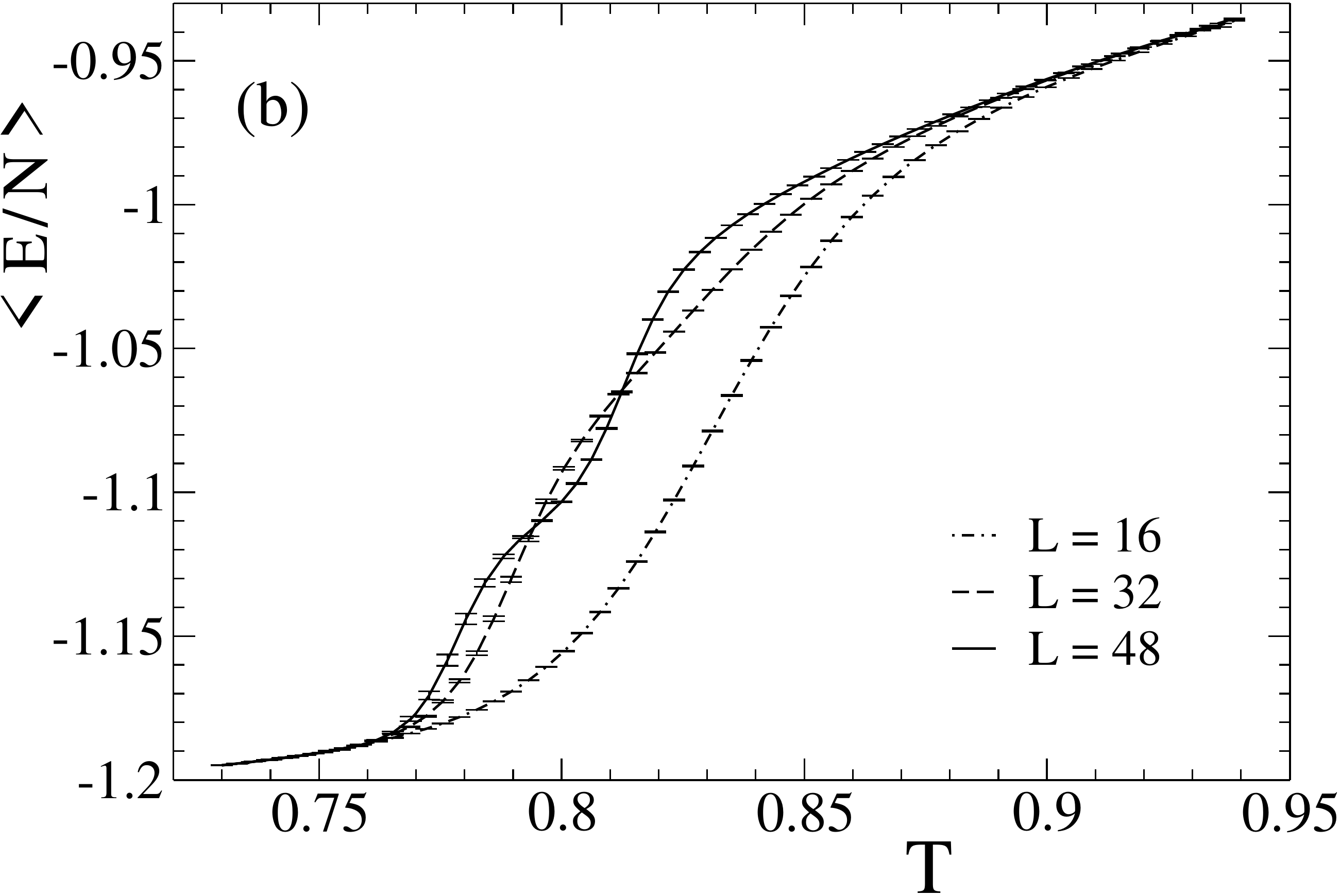}}
\end{minipage}

\vspace{0.8cm}

\begin{minipage}[h]{0.50\textwidth}
\subfigure{\includegraphics[width=0.95\textwidth]{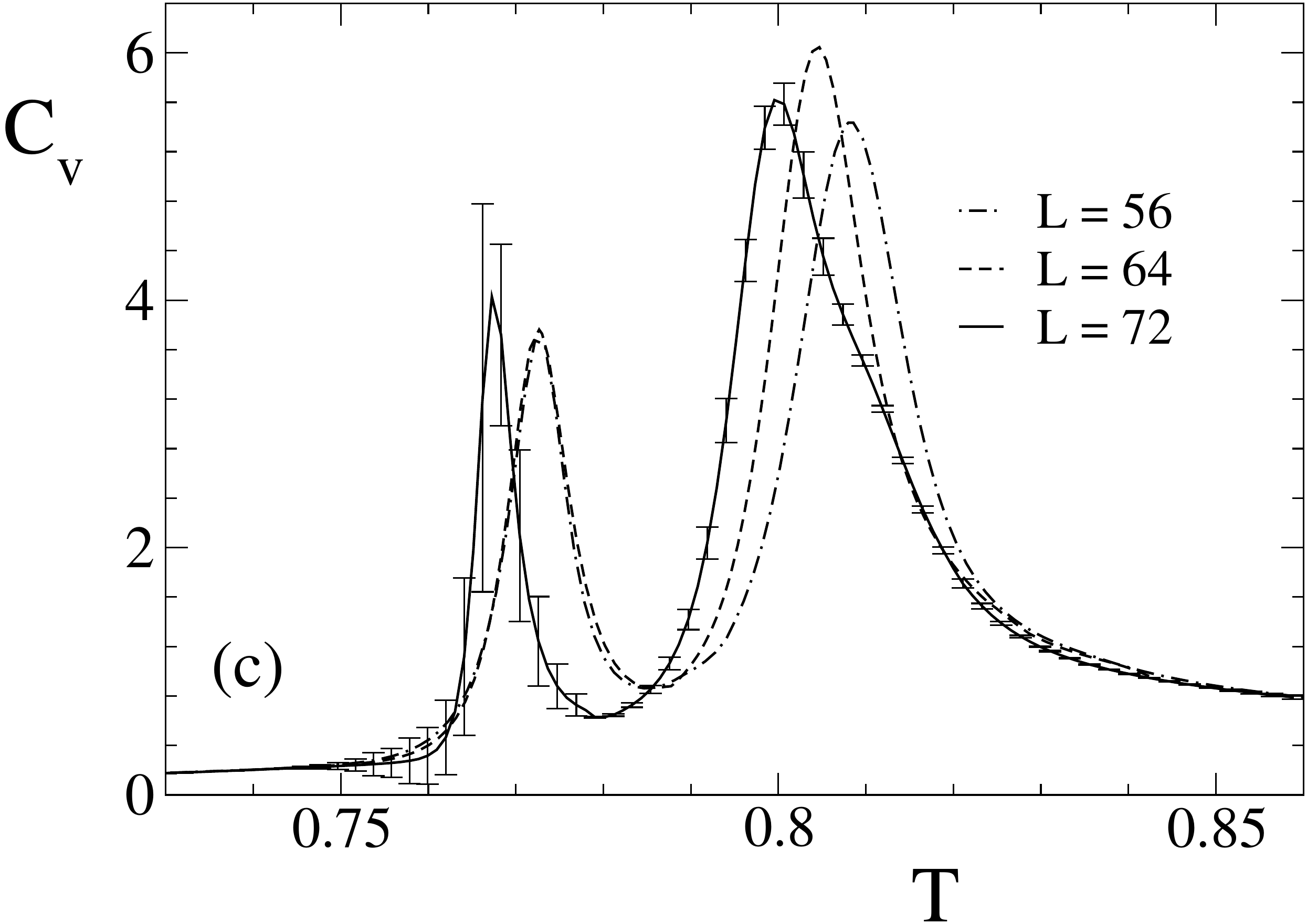}}
\end{minipage}%
\begin{minipage}[h]{0.50\textwidth}
\subfigure{\includegraphics[width=0.95\textwidth]{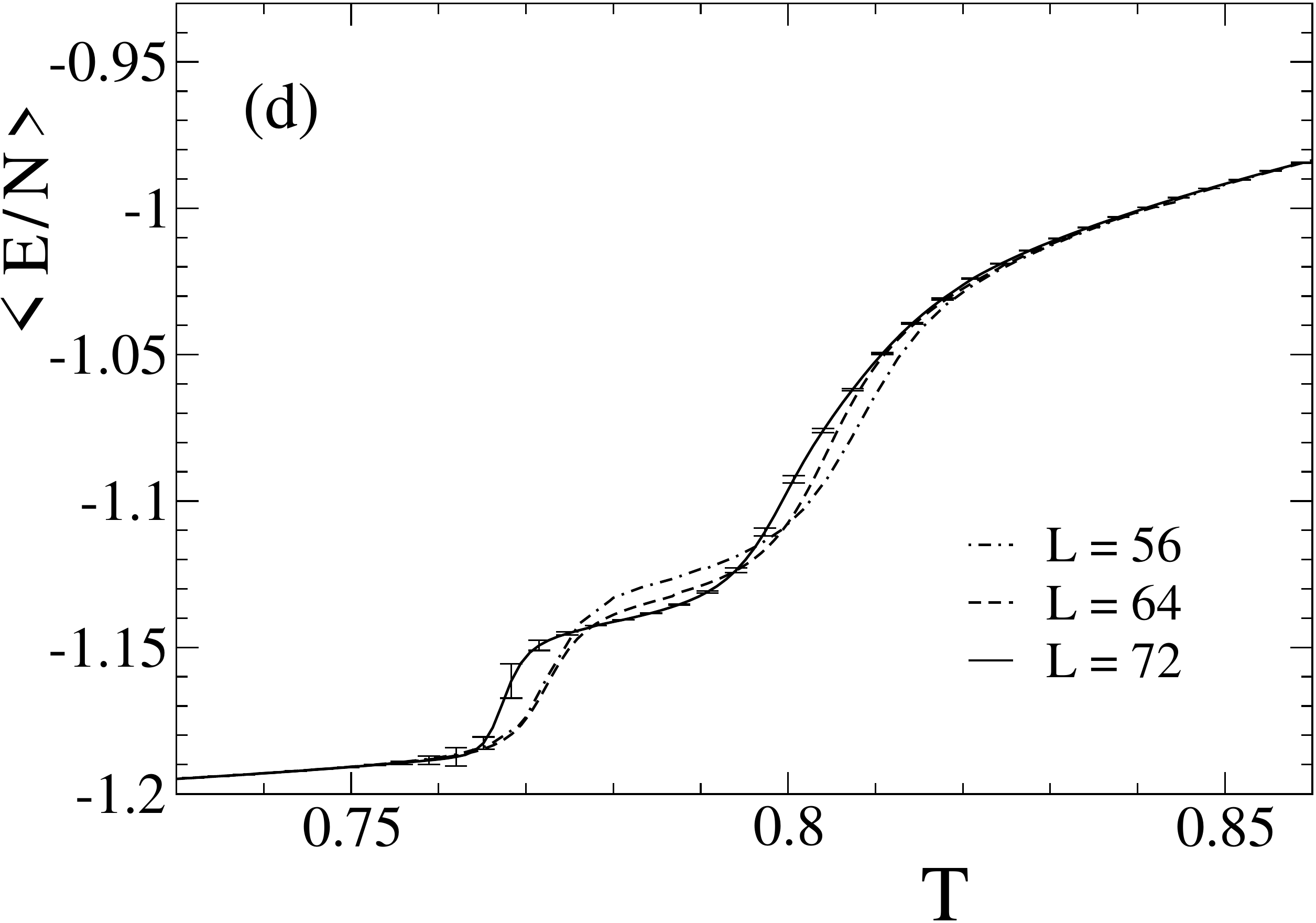}}
\end{minipage}
\end{center}
\caption{ Specific heat $C_v$ and average energy per spin $\langle E/N \rangle$ as a function of temperature $T$ 
are shown in figures (a) and (b) for $L=16, 32, 48$, and in figures (c) and (d) for $L=56, 64, 72$. 
  The inset in figure 1(a) shows $C_v$ with error bar estimates for $L=48$.}
\label{fig:1}
\end{figure}

  To increase the statistics for the final reweighting, we combine data obtained from independent simulations at temperatures $T_0$, as shown in table 1, with the condition that the $P$ simulations occur at $T_0$ values close to each other \cite{alves_376_1990}.
  This technique is based on the remark that the final estimate $\overline{f(T)}$,
obtained from $P$ independent statistical quantities $\overline{f_i}$,
can be calculated as a weighted linear combination
\begin{equation}
        \overline{f(T)} =  \sum_{i=1}^{P}  w_i\, \overline{f_i(T)} \, ,
\end{equation}
  where $w_i = w_i(T)$ are normalized weights, which one takes as the inverse statistical variance 
$w_i(T) \approx  1/\sigma^2(\overline{f_i})$ from each MC time series.
  The overall constant is determined by the normalization condition $\sum_{i=1}^P w_i=1$.
  This procedure corresponds to increasing the final statistics $P$ times.
  

\begin{figure}[!t]
\begin{center}
\begin{minipage}[h]{0.27\textwidth}
\subfigure[$~~T = 0.710$]{\includegraphics[width=0.8\textwidth]{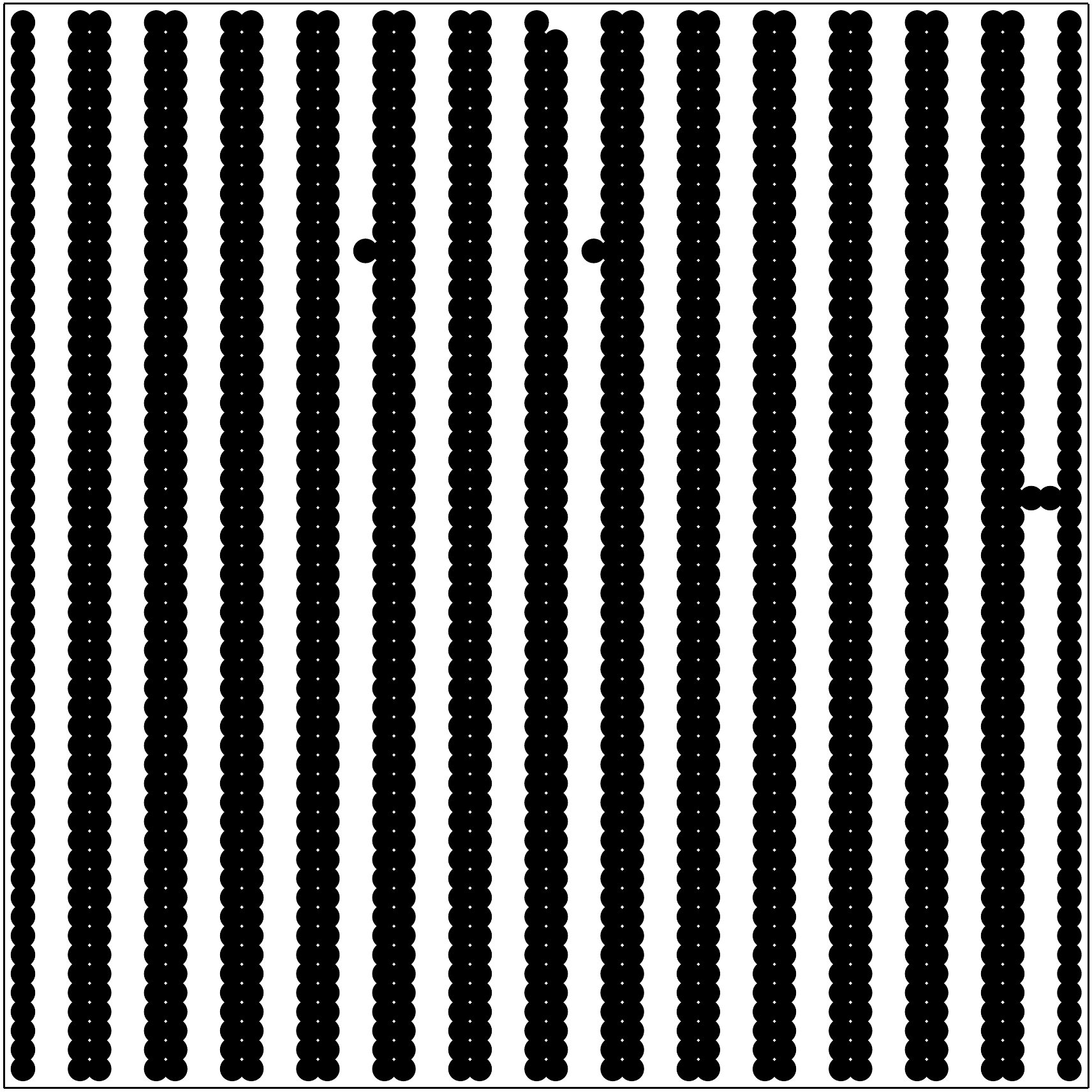}}
\end{minipage}%
\begin{minipage}[h]{0.27\textwidth}
\subfigure[$~~T = 0.790$]{\includegraphics[width=0.8\textwidth]{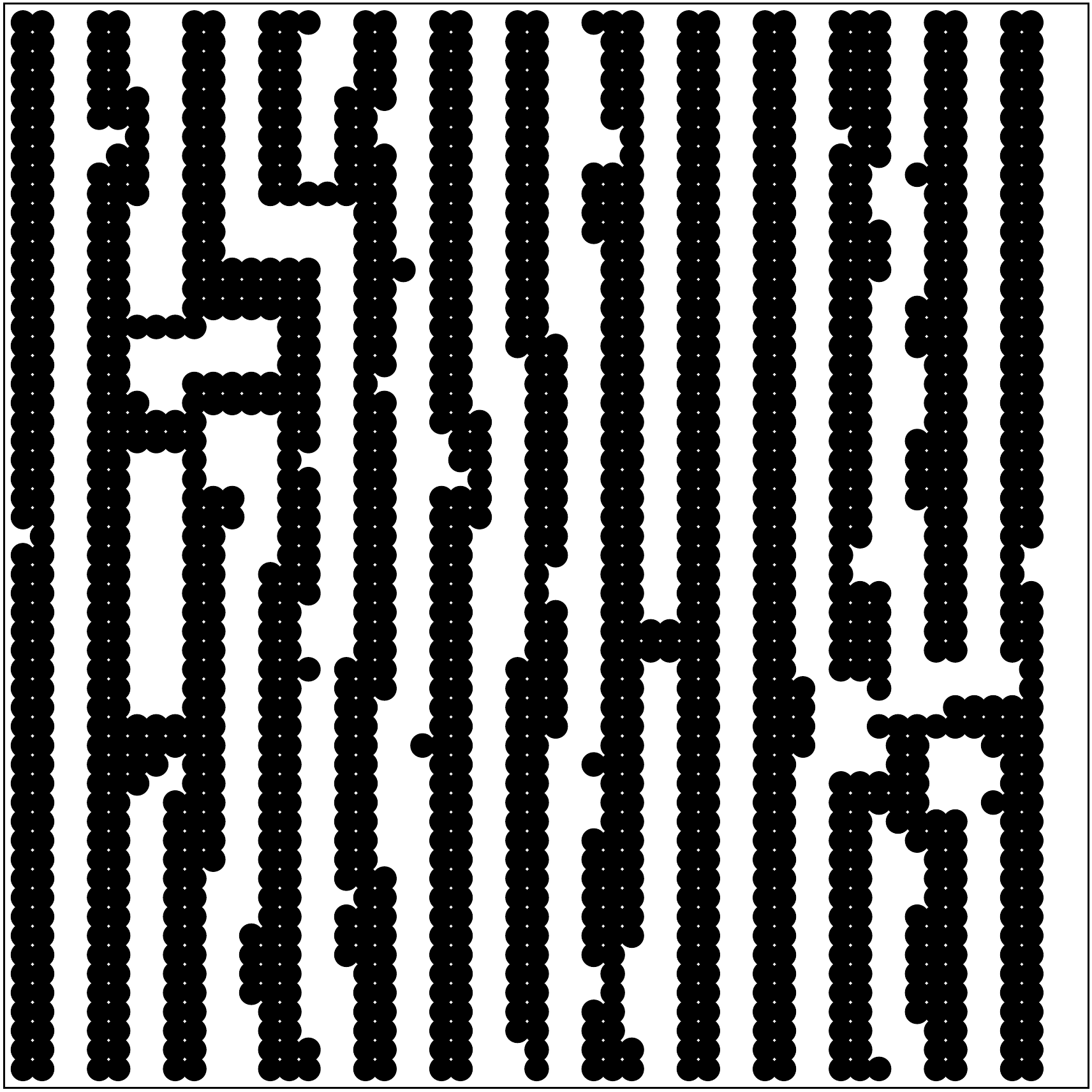}}
\end{minipage}
\begin{minipage}[h]{0.27\textwidth}
\subfigure[$~~T = 0.840$]{\includegraphics[width=0.8\textwidth]{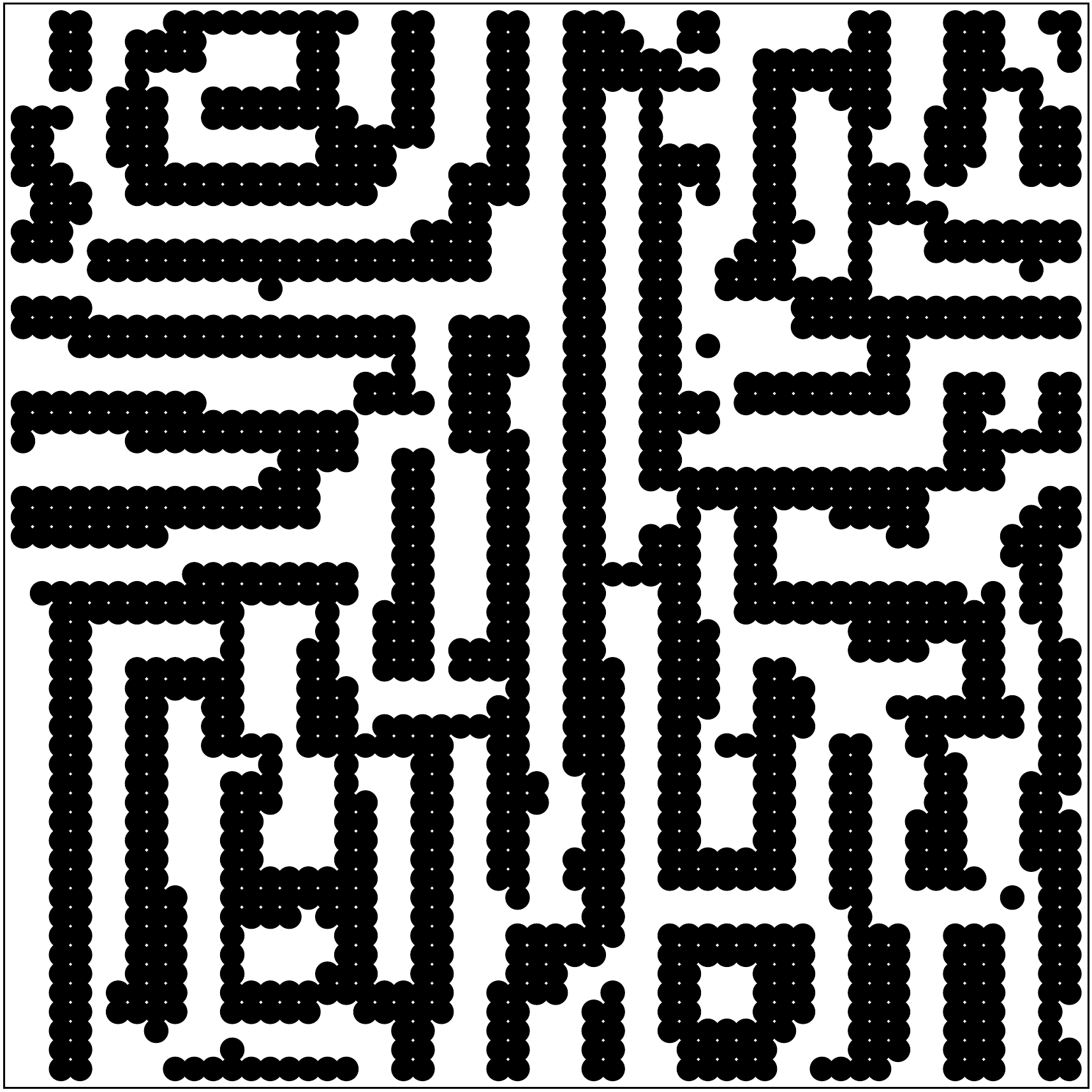}}
\end{minipage}
\caption{Typical spin configurations for $L=56$ at temperatures in the: (a) striped, (b) nematic and (c) tetragonal phases. The configuration displayed in the striped phase presents $E/N=-1.2028$, $O_{hv}=0.9873$; in the nematic phase, $E/N=-1.1172$, $O_{hv}=0.7903$; and in the tetragonal liquid phase $E/N=-1.0201$ and $O_{hv}=0.1777$.}
\label{fig:2}
\end{center}
\end{figure}

\begin{figure}[!b]
\begin{center}
\begin{minipage}[h]{0.27\textwidth}
\subfigure[$~~T = 0.773$]{\includegraphics[width=0.8\textwidth]{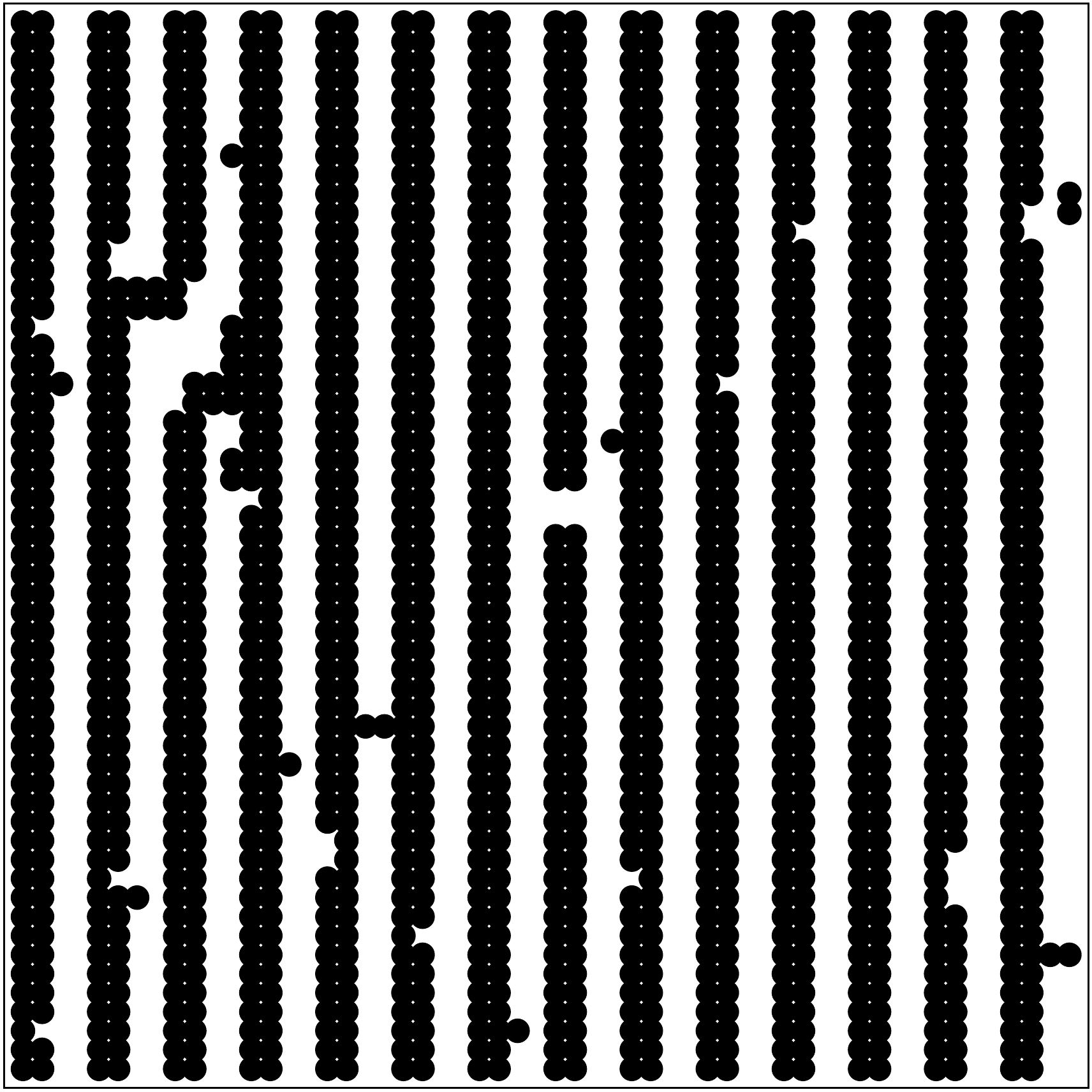}}
\end{minipage}%
\begin{minipage}[h]{0.27\textwidth}
\subfigure[$~~T = 0.808$]{\includegraphics[width=0.8\textwidth]{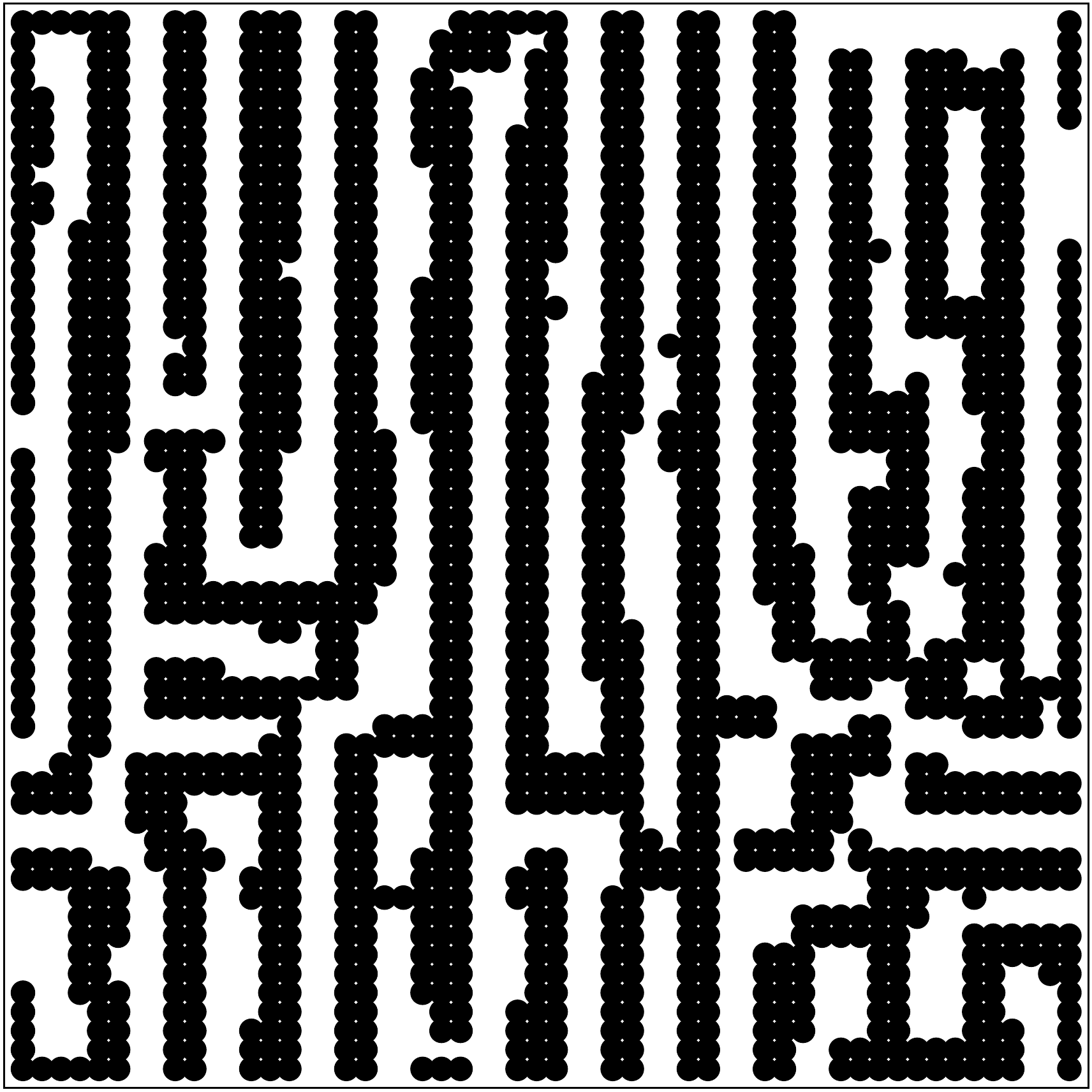}}
\end{minipage}
\caption{Spin configurations at the transition temperatures for $L=56$.
The configuration displayed in (a) has $ E/N =-1.1670$, $ O_{hv} =0.9205$;
while in (b), $ E/N =-1.0769$ and $O_{hv}=0.5633$.}
\label{fig:3}
\end{center}
\end{figure}


  Figure 1 shows our final estimates for the averages of specific heat $C_v$ and energy per spin  
$\langle E/N \rangle$ as a function of temperature.
  Figure 1(a) displays $C_v$ for $L=16, 32$, and 48.
  We have included error bar estimates for $L=48$ only in the inset of Fig. 1(a), 
to have a clear presentation of the $C_v$ behavior for different lattice sizes.  
  The numerical simulation with $L=16$ shows a typical striped-tetragonal transition characterized
by a peak in $C_v$.
  This behavior changes dramatically and gives origin to an intermediate thermodynamic transition
as we increase the lattice size.
  Moreover, this new transition, the nematic-tetragonal transition, presents a more pronounced 
increase in the specific heat maximum compared with that of the striped-nematic transition,
which occurs at a lower temperature.

   Figure 2 illustrates the typical spin configurations in each phase for $L=56$.
   This figure shows how the directional and translational symmetries are missed
as we increase the temperature toward the tetragonal phase. 
   The corresponding typical spin configurations at the 
transition temperatures are displayed in figure 3.

  The finite-size critical temperatures $T_c(L) = T_{\rm max}(L)$ are defined by the maximum of the specific heat $C_v(L)$.
  Making Jackknife bins from the combined MC statistics, we reweight 
each bin in order to find the maximum of the specific heat. 
  This procedure leads to final estimates of critical temperatures and their error bars in table 1.
 

 We note that the specific heat peaks occur at lower temperatures as $L$ increases.
  This is another finite-size effect.
  In fact, finite-size scaling (FSS) arguments predict the following behavior for the shift in the critical temperatures
\begin{equation}
 T_c(L)- T_c(\infty) \sim L^{-1/ \nu} \, ,
\end{equation}
where $\nu$ is the critical exponent.
  This exponent assumes the value $1/d$ for first-order phase transitions, where $d$ is the system dimension.
  Another scaling relation for the shift in the critical temperatures
can be considered in the case of a Kosterlitz-Thouless (KT) type transition \cite{monousakis_94},
\begin{equation}
 T_c(L)- T_c(\infty) \sim  \frac{a}{(\log L - b)^2} \, ,
\end{equation} 
where $a$ and $b$ are constants.
  These are multiparameter fits, which do not allow us to estimate the infinite volume critical temperature
$T_c(\infty)$ or the proper scenario.
   In the case of a KT type transition, the specific heat peak does not occur at the
critical temperature but above the transition temperature \cite{van_1981}.

   From the results in table 1 for the finite-size critical temperatures, and because we expect to observe both phase transitions also in the thermodynamic limit, it is reasonable to suppose that the difference in the critical temperatures will reach a fixed value.

   To further characterize the observed sequence of phases, the degree of orientational order
   \cite{Rastelli_PRB76_2007,ibooth_75_1995},
\begin{equation}
O_{hv} = \left| \frac{n_{v}-n_{h}}{n_{v}+n_{h}} \right|,
\end{equation}
is also computed for each configuration of the system.
  The quantities $n_{h}$ and $n_{v}$ denote the number of horizontal and vertical bonds between oppositely aligned nearest-neighbor spins.
  For the striped ground state, this order parameter takes the value +1, while for higher temperatures, in which the orientational symmetry of the striped domains is broken, it vanishes.
  To see how this order parameter and its susceptibility
\begin{equation}
\chi(O_{hv}) = N \left( \langle O_{hv}^{2} \rangle  -  \langle O_{hv} \rangle^{2} \right),
\end{equation}
characterize the striped, nematic, and tetragonal phases, we also study their behavior as a function of
temperature by means of reweighting techniques.

We present the maxima of the susceptibilities and their respective 
finite-size critical temperatures $T_{\rm max}$ in table 2.
  These temperatures are defined by the maximum of susceptibility $\chi(T)$
as shown in figure 4(a) and (c).
  Figures 4(b) and (d) illustrate the behavior of the order parameter $O_{hv}$ as a function of temperature for different lattice sizes.
  They show that the break of the directional symmetry happens mainly at the second thermodynamic transition temperature $T_c^{(2)} \approx 0.809$.
  The peaks in the susceptibility related to this second transition become higher as
the lattice size increases.
  On the other hand, it seems we have a trend of lower values for its maximum as the
lattice size increases in the case of the first thermodynamic transition temperature 
$T_c^{(1)} \approx 0.77$.
  According to the estimates in table 2 and their error bars, we are facing an unusual trend of decreasing peaks that
 continues until a limiting value is reached.
  In any case, this new feature places an extra degree of difficulty to the identication of the character of the transition.
  This situation presents a parallel behavior when the maxima of the specific heat is analyzed 
in table 1.
  The first transition ($T_c^{(1)} \approx 0.77$) seems to present specific heat peaks of roughly the same order of magnitude, while the second transition now located at $T_c^{(2)} \approx 0.800$ shows a trend of higher peaks as the lattice size increases.

\begin{table}
\begin{center}
\begin{tabular}{@{}lllllll} 
\hline \hline
~$L$ & ~& $\chi(O_{hv})_{\rm max}$    & ~~$T_{\rm max}^{(1)}$ 
     &~~~~& $\chi(O_{hv})_{\rm max}$    & ~~$T_{\rm max}^{(2)}$ \\
\hline
 ~16 & ~&          &          &~~~~& 28.42(2) & 0.8336(2) \\
 ~32 & ~&          &          &~~~~&  49.4(1) & 0.8258(2) \\
 ~48 & ~&  17.0(5) & 0.781(2) &~~~~& 129.7(6) & 0.8140(2)  \\
 ~56 & ~&  13.8(2) & 0.773(1) &~~~~& 170(2)   & 0.8097(2)  \\
 ~64 & ~&  13.5(3) & 0.773(1) &~~~~& 191(2)   & 0.8073(5)  \\
 ~72 & ~&  13.1(9) & 0.768(2) &~~~~& 209(3)   & 0.809(2)  \\
\hline \hline
\end{tabular}
\caption{Susceptibility maxima and the corresponding critical temperatures $T_{\rm max}$ 
as shown in figure 4.}
\end{center}
\end{table}

\begin{figure}[!th]
\begin{center}
\begin{minipage}[h]{0.50\textwidth}
\subfigure{\includegraphics[width=0.95\textwidth]{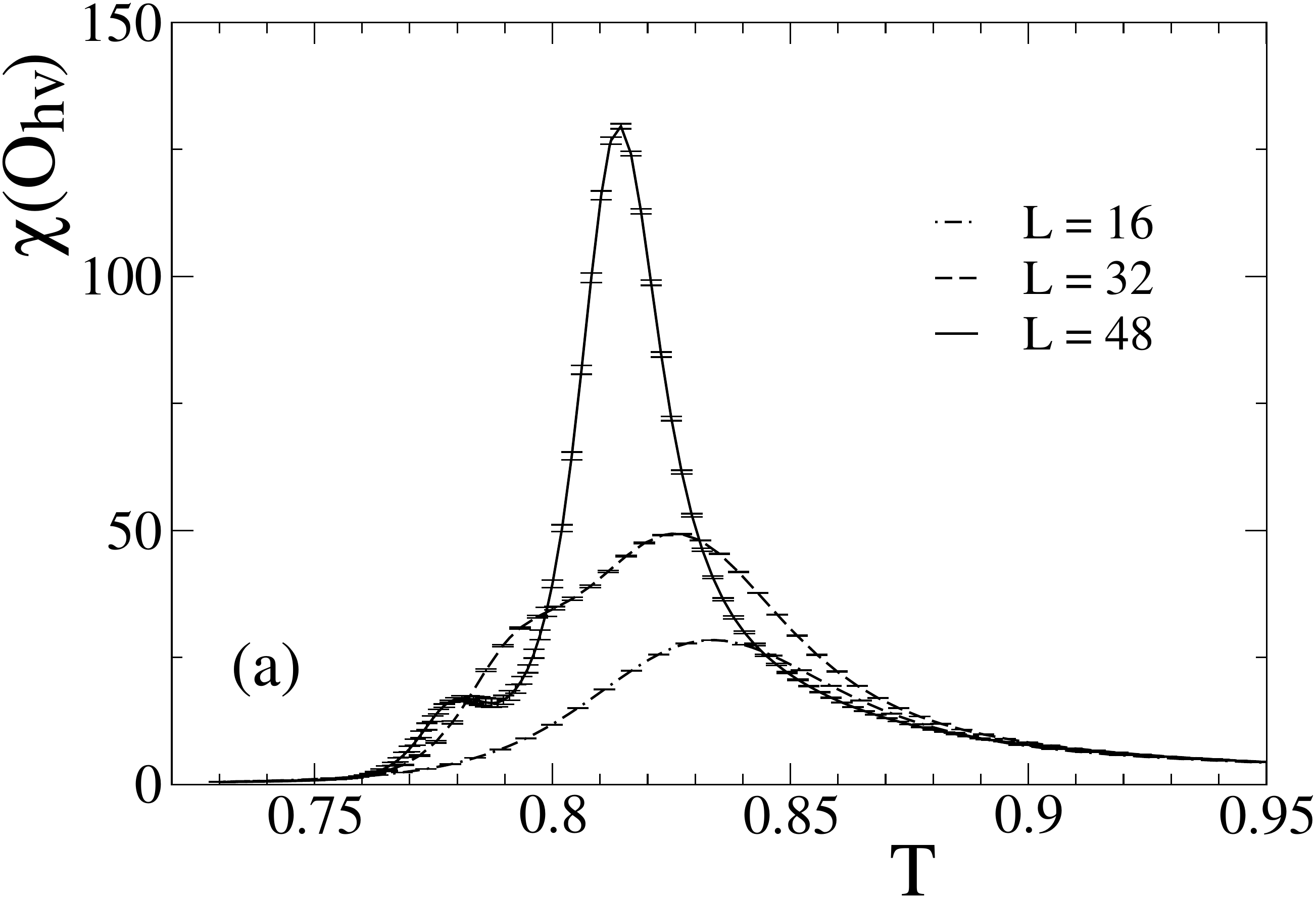}}
\end{minipage}%
\begin{minipage}[h]{0.50\textwidth}
\subfigure{\includegraphics[width=0.95\textwidth]{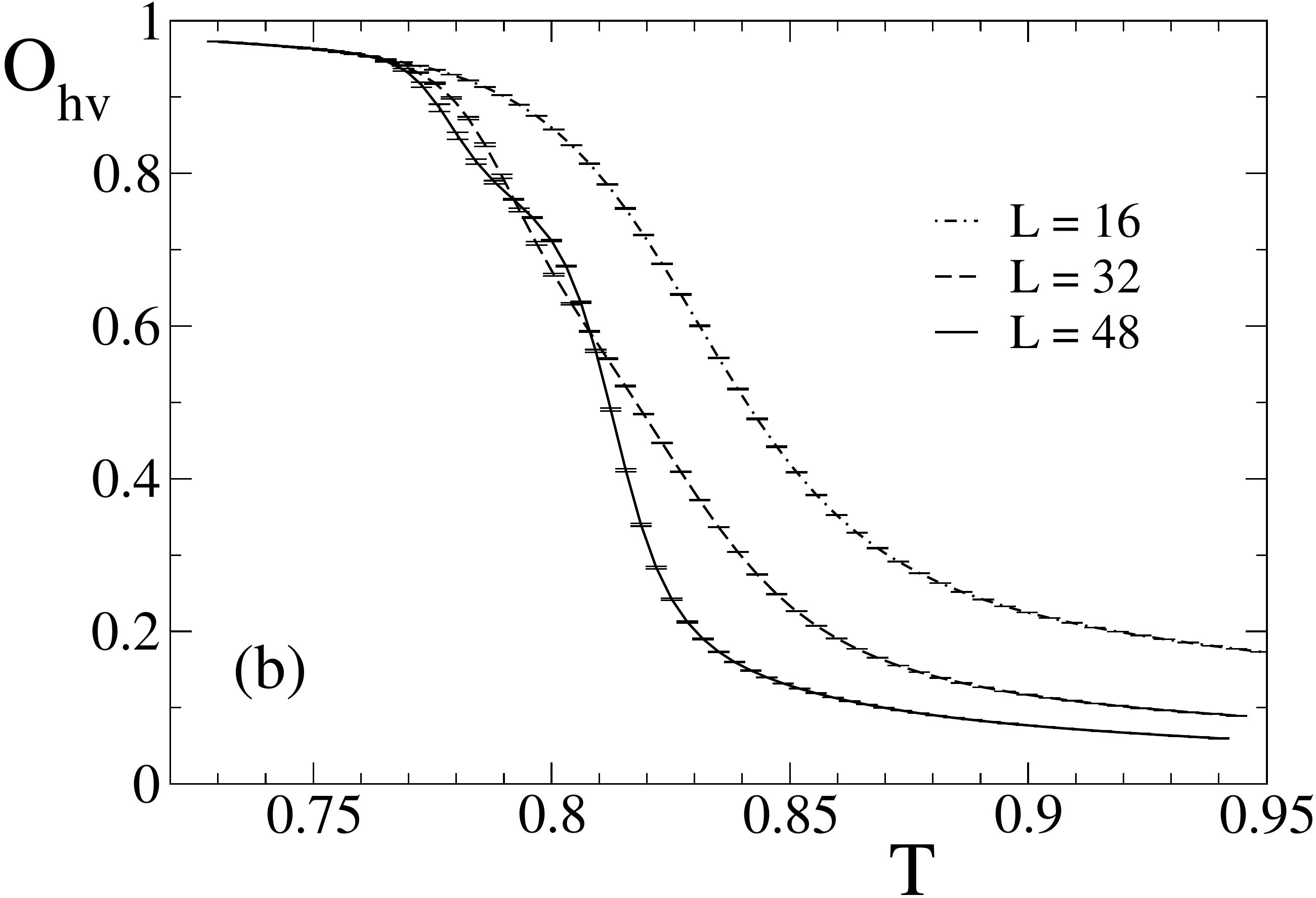}}
\end{minipage}

\vspace{0.8cm}

\begin{minipage}[h]{0.50\textwidth}
\subfigure{\includegraphics[width=0.95\textwidth]{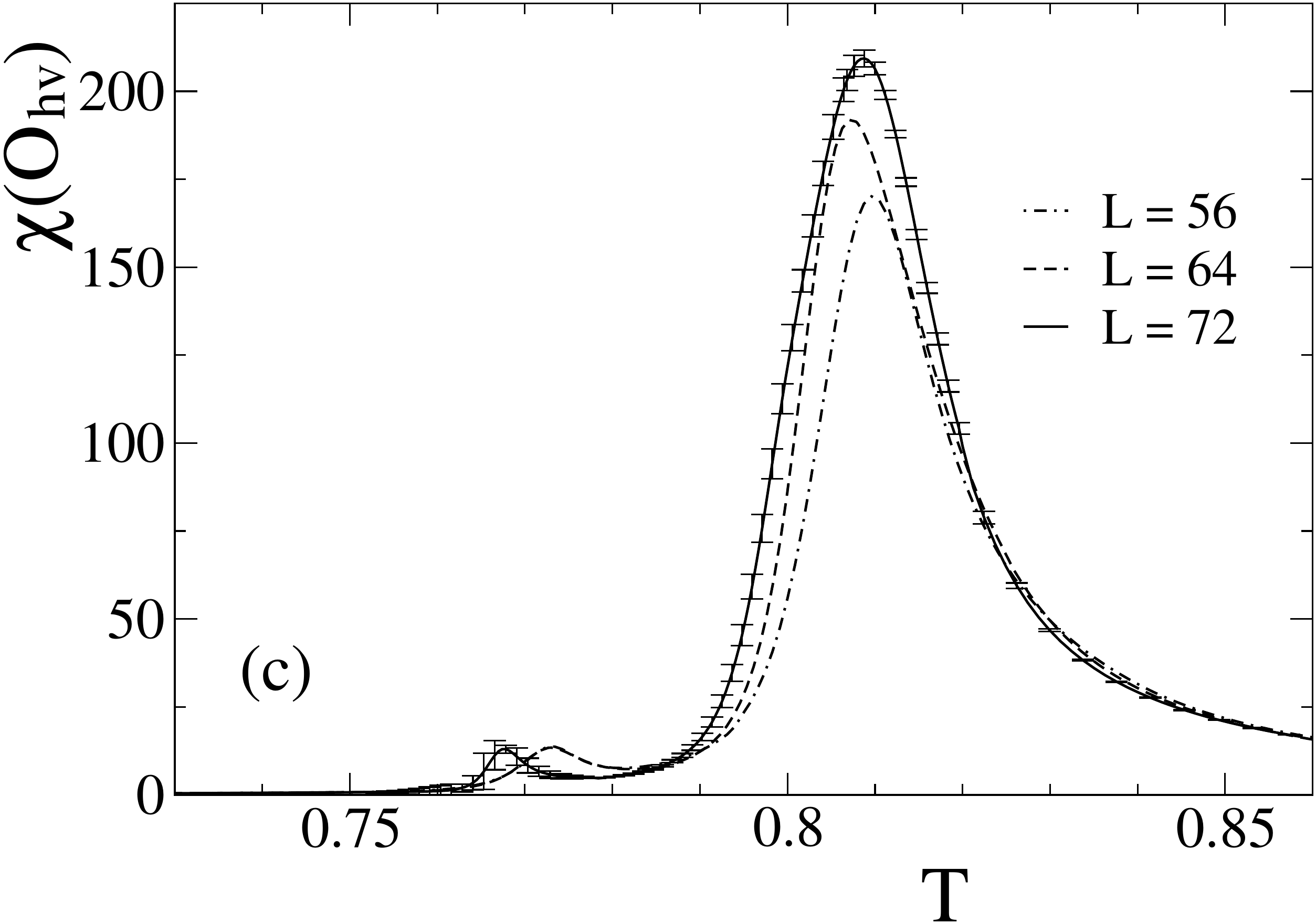}}
\end{minipage}%
\begin{minipage}[h]{0.50\textwidth}
\subfigure{\includegraphics[width=0.95\textwidth]{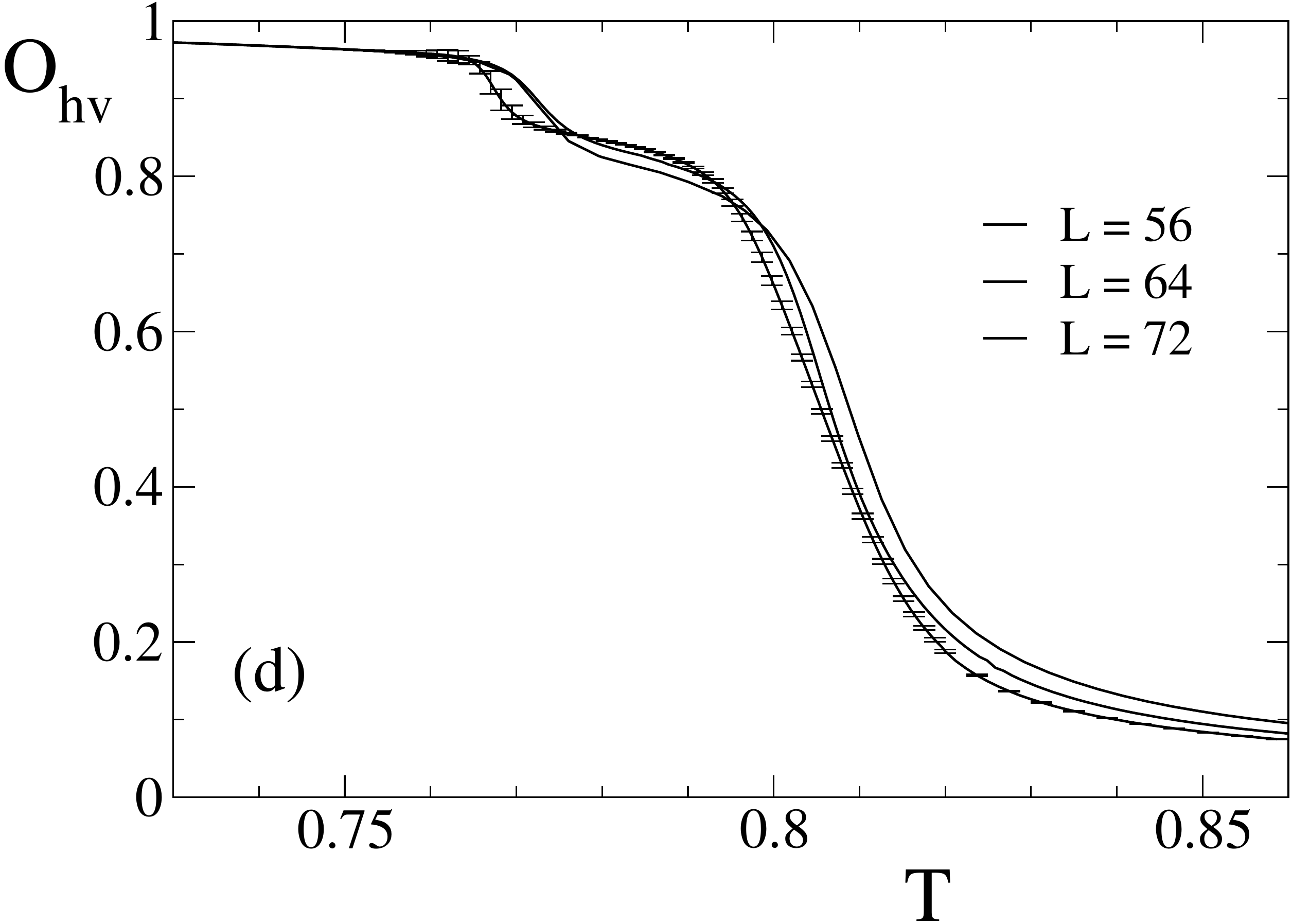}}
\end{minipage}
\end{center}
\caption{Susceptibility and order parameter for lattice sizes $L=16 - 72$ as a function of temperature.}
\label{fig:4}
\end{figure}

   The characterization of the order of the observed thermodynamic phase transitions can be achieved by means of FSS analysis.
   In this context, the specific heat peak is described by the relation,
\begin{equation}
    C_{v}|_{\rm max} \sim N^{\alpha/d \nu} \, .
\end{equation}
   The FSS relation for the peak in the susceptibility is given by
\begin{equation}
   \chi_{\rm max} \sim N^{\gamma/d \nu} \, .
\end{equation} 
   The first-order character is related to the critical exponent $d\nu = 1$ \cite{Fisher_1982,Decker_1988}.
   From the hyperscaling relation $\alpha = 2 -d\nu$, and assuming a first-order transition,
one obtains $\alpha/d \nu = 1$ and $\gamma/d \nu =1$.

   Since the phenomenology of this model reveals itself only for rather large lattice sizes, our data
still present limitations for a reliable FSS analysis.
   As we are going to show in the next Section, any intensive MC simulations with local update algorithms
will suffer severe limitations.
   However, our data allows one to draw some conclusions.
   
   First of all, let us verify the possible scenario found by the FSS analysis.
   To this end, we must only consider the last four results for the maximum of $C_v$ in table 1
and the maximum of $\chi(O_{hv})$ in table 2,  corresponding to $L=48, 56, 64 $ and 72.
   Since the peaks for both specific heat and susceptibility related to the first transition
seem to converge toward a constant value, we may assume $\alpha \approx 0$ and $ \gamma \approx 0$.
   In this case, logarithmic corrections should be taken into account to describe their
maximum.
   The supposed values for the critical exponents do not support the first-order transition character.
   If $C_v$ does not present a divergent behavior, nor has appreciable finite-size effects,
then the hypothesis of a Kosterlitz-Thouless type transition must be considered \cite{abanov_54_1995}.
   The hypothesis of a KT type transition has been discussed as an alternative to the possibility
of the first-order character found from MC studies in Ref. \cite{cannas_75_2007,cannas_73_2006}.
   Here, in spite of the high precision values for the physical estimates,  
our results for the temperature and response functions may be biased when one considers estimates from the two largest lattice sizes, mainly the $L=72$ case, hampering further conclusions. 
   In particular, the lower precision in the specific heat at $T_c^{(1)}$ when compared with the results for the second peak at $T_c^{(2)}$ is related to strong autocorrelations in the striped-nematic transition, as presented in the next section.
        
   The FSS analysis for the second transition yields
$\alpha = 0.80(2)$, $\gamma = 0.77(4)$ and $d\nu = 1.20(2)$ for $L=48 -72$.
   These results clearly support the second order character for the nematic-tetragonal phase transition
as first suggested theoretically by Abanov \cite{abanov_54_1995}, via the continuous model,
and by Pigh\'in and Cannas \cite{cannas_75_2007} from a mean field approach.
   However, we are uneasy about these numerical findings because of their very small godness-of-fit.
   For $L=72$, in particular, we may argue that $C_v$ evaluation presents significant bias since its value 
does not follow the main trend, leading the scaling relation to an unlikely fit. 
   However, this picture does not change if we discard $L=72$. 
   We obtain roughly the same numerical estimates for the exponents because $C_v(L=72)$ has comparatively large error bar.

   Now, we will analyse the energy density distributions and, as explained below, we include the calculation of free energies and interface tensions, in our case linear
tensions, to gather further information about the possible nature of the phase transitions.

\begin{figure}[h]
\begin{center}
\begin{minipage}[h]{0.50\textwidth}
\subfigure{\includegraphics[width=0.95\textwidth]{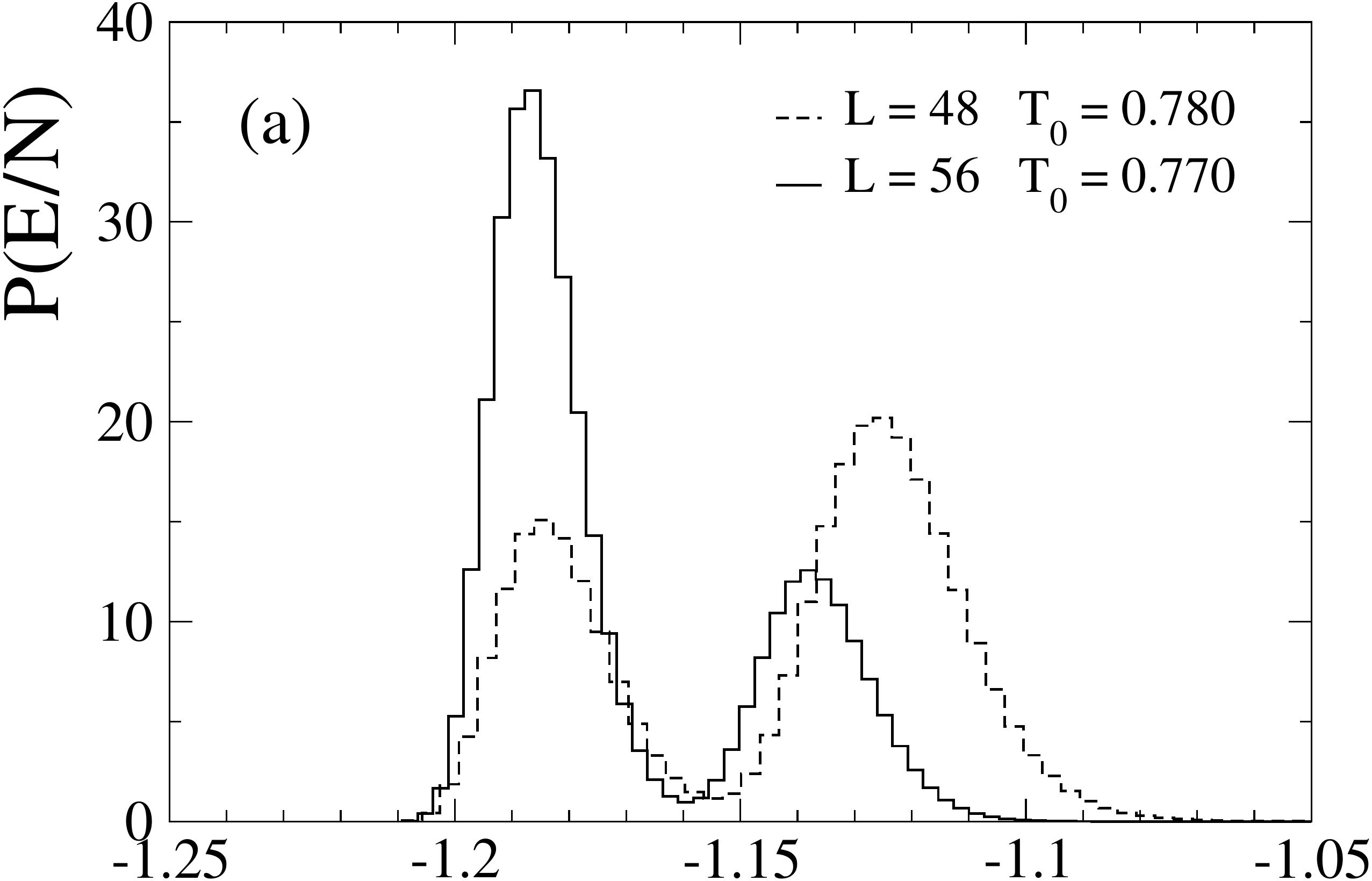}}
\end{minipage}%
\begin{minipage}[h]{0.50\textwidth}
\subfigure{\includegraphics[width=0.90\textwidth]{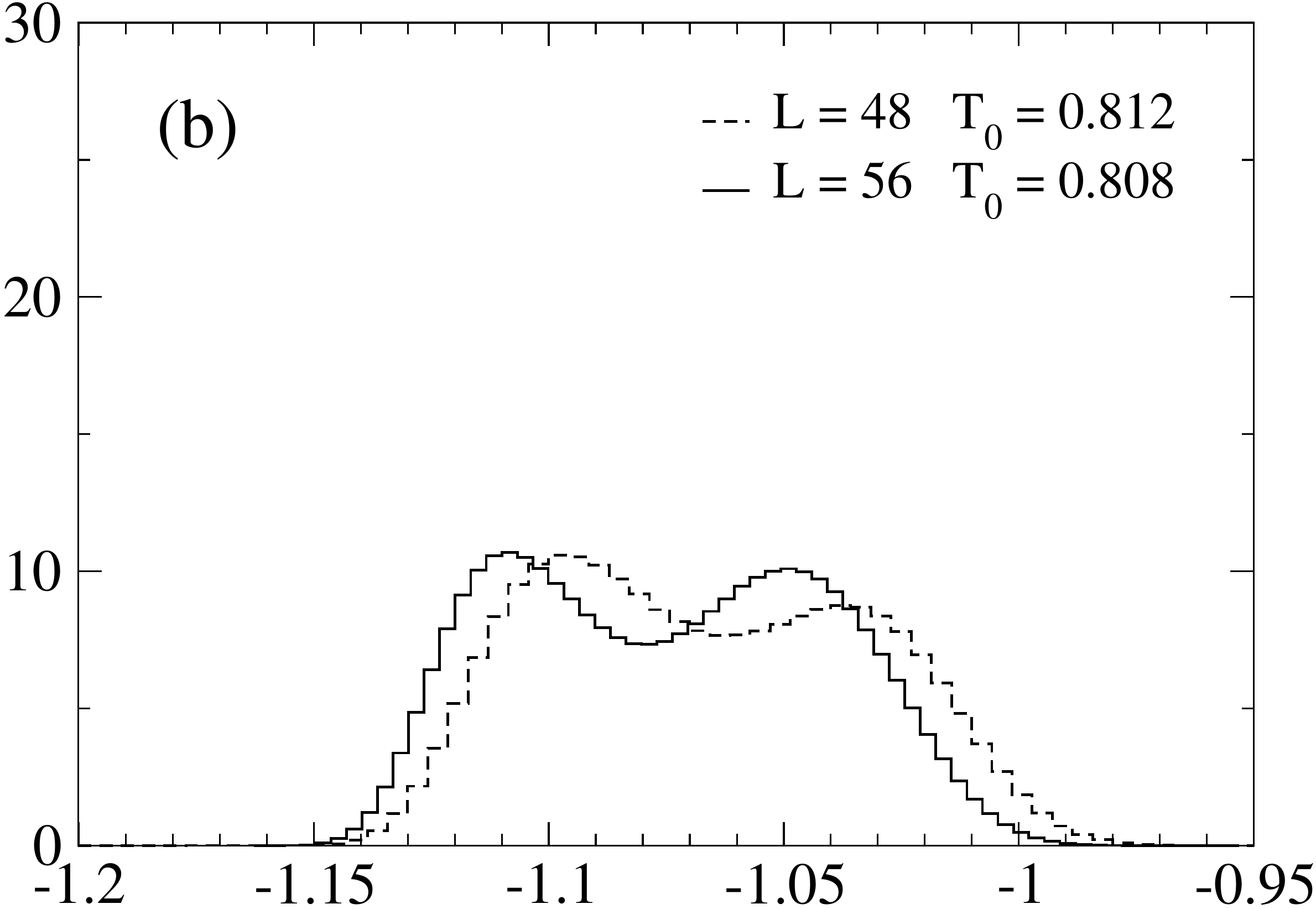}}
\end{minipage}

\vspace{0.8cm}

\begin{minipage}[h]{0.50\textwidth}
\subfigure{\includegraphics[width=0.95\textwidth]{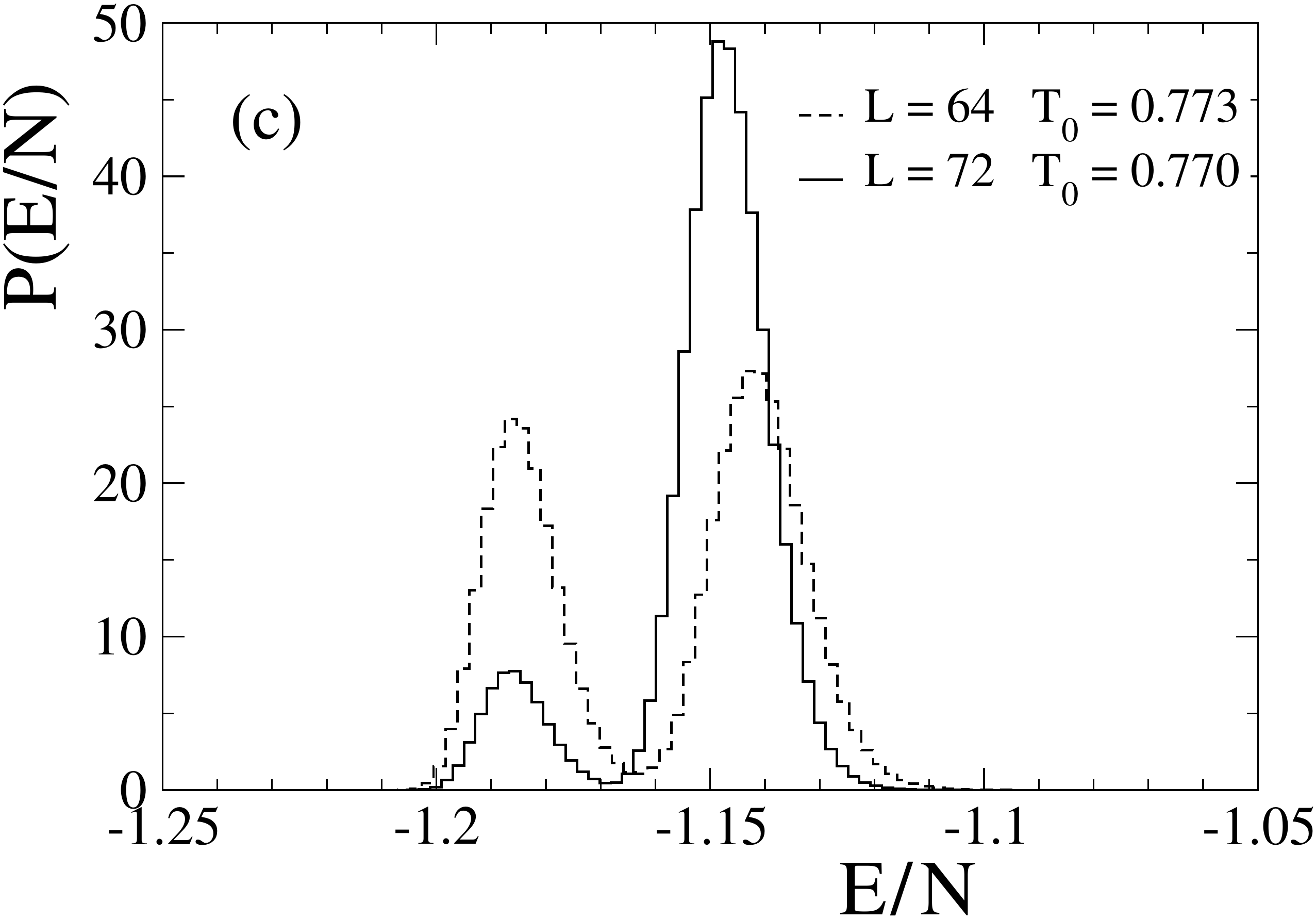}}
\end{minipage}%
\begin{minipage}[h]{0.50\textwidth}
\subfigure{\includegraphics[width=0.90\textwidth]{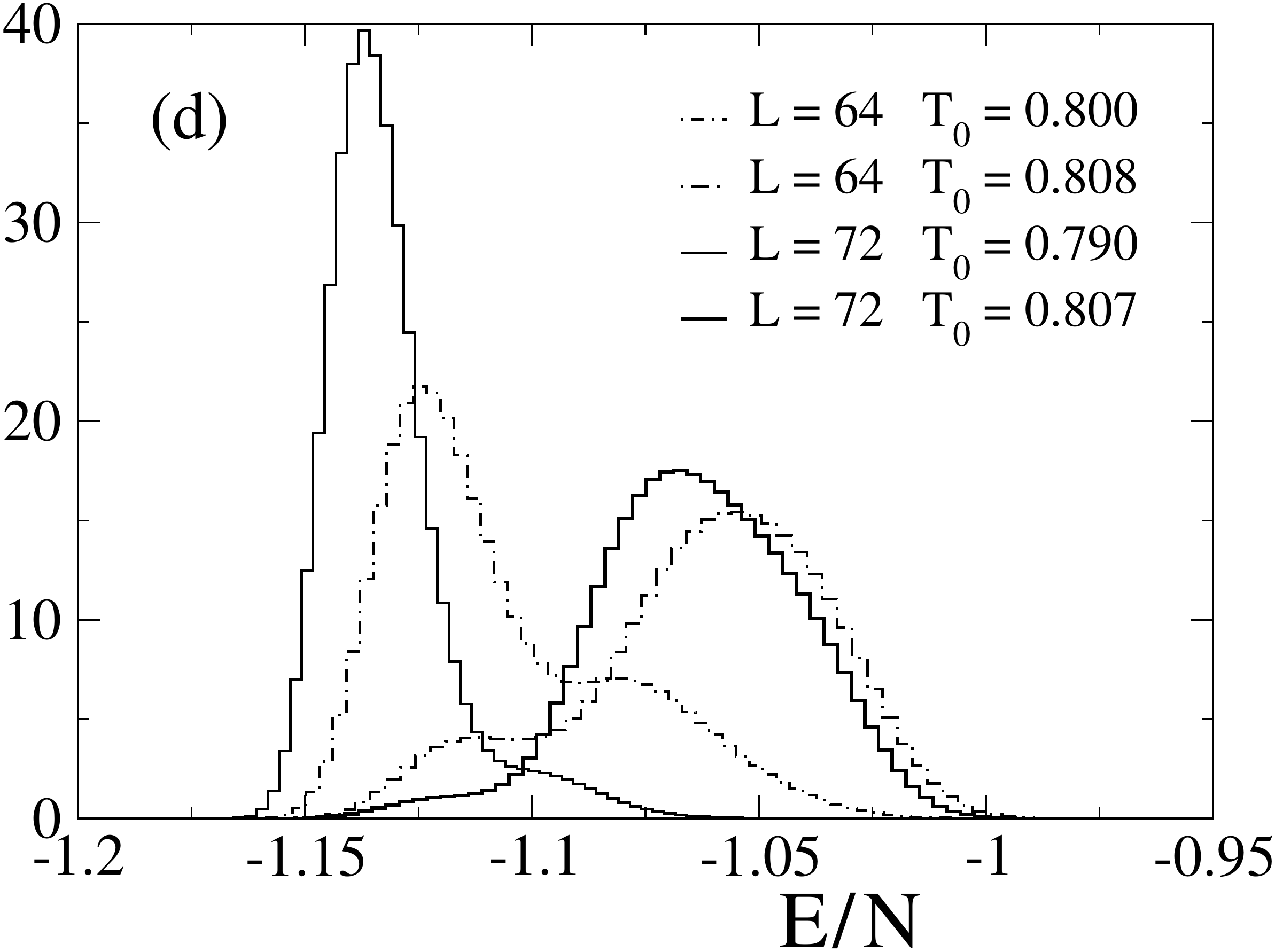}}
\end{minipage}
\end{center}
\caption{Energy density distributions for lattice sizes $L=48 - 72$ from simulations at temperatures $T_0$.}
\label{fig:5}
\end{figure}

   In Fig. 5 we show energy density distributions from data obtained at temperatures $T_0$.
   Here, we exhibit only energy distributions from simulations at temperatures close to the critical ones.
   In Fig. 5(d) we include two distributions for each lattice size because none of the sampled temperatures are close enough to the critical ones to display their true critical distributions.
   Figure 6 shows all histograms of energy distributions for $L=64$. 	
  
   Let us initially consider the first transition.
   Figures 5(a) and (c) present histograms with clear double-peak structures close to the critical temperatures $T_c^{(1)}$.
   This is a typical first-order phase transition behavior where a latent heat is responsible for the existence of
those peaks at energies $E_{st}(L)$ and $E_{nem}(L)$, corresponding to the striped and nematic phases, respectively. 
   Those states are separated by a minimum at $E_m(L)$ corresponding to domains describing the coexistence of both phases.
   Here, we can evaluate the free energy barrier $\Delta F(L)$ for this temperature-driven transition.
   Although it seems clear we face a first-order transition,  it remains to be seen whether the linear tension 
is not negligible in the thermodynamic limit.
   To this end, we evaluate the free energy \cite{binder_1982,kosterlitz_1991}
\begin{equation}
 \Delta F(L) = \ln \frac{P_L(E_{st}/N,T_h)}{P_L(E_m/N,T_h)} 
\end{equation}
at $T_h$, where the finite lattice critical point $T=T_h(L)$ is defined such that 
the histogram peaks have equal heights $ P_L(E_{st}/N,T_h) = P_L(E_{nem}/N,T_h)$.
   We proceed performing patching and reweighting of histograms \cite{alves_1992},
in order to evaluate $\Delta F(L)$ at both transitions.
  This leads to the results in Fig. 7 for both transitions.
  In this figure, the solid and dashed lines are the linear extrapolation in $1/L$,
\begin{equation}
 \frac{\Delta F(L)}{2L} = f_0 + \frac{f_1}{L} \, ,
\end{equation}
where $f_0$ is the linear tension and $f_1$ stands for possible FSS correction.
  Here, we assume the two-gaussian approximation \cite{binder_1982,binder_1986}
to describe the coexisting phases in this model.
  This procedure yields  $f_0 = 0.019(4)$  and $f_1= 0.37(16)$ for the striped-nematic transition.

\begin{figure}[!t]
\begin{center}
\begin{minipage}[h]{0.70\textwidth}
\includegraphics[width=0.8\textwidth]{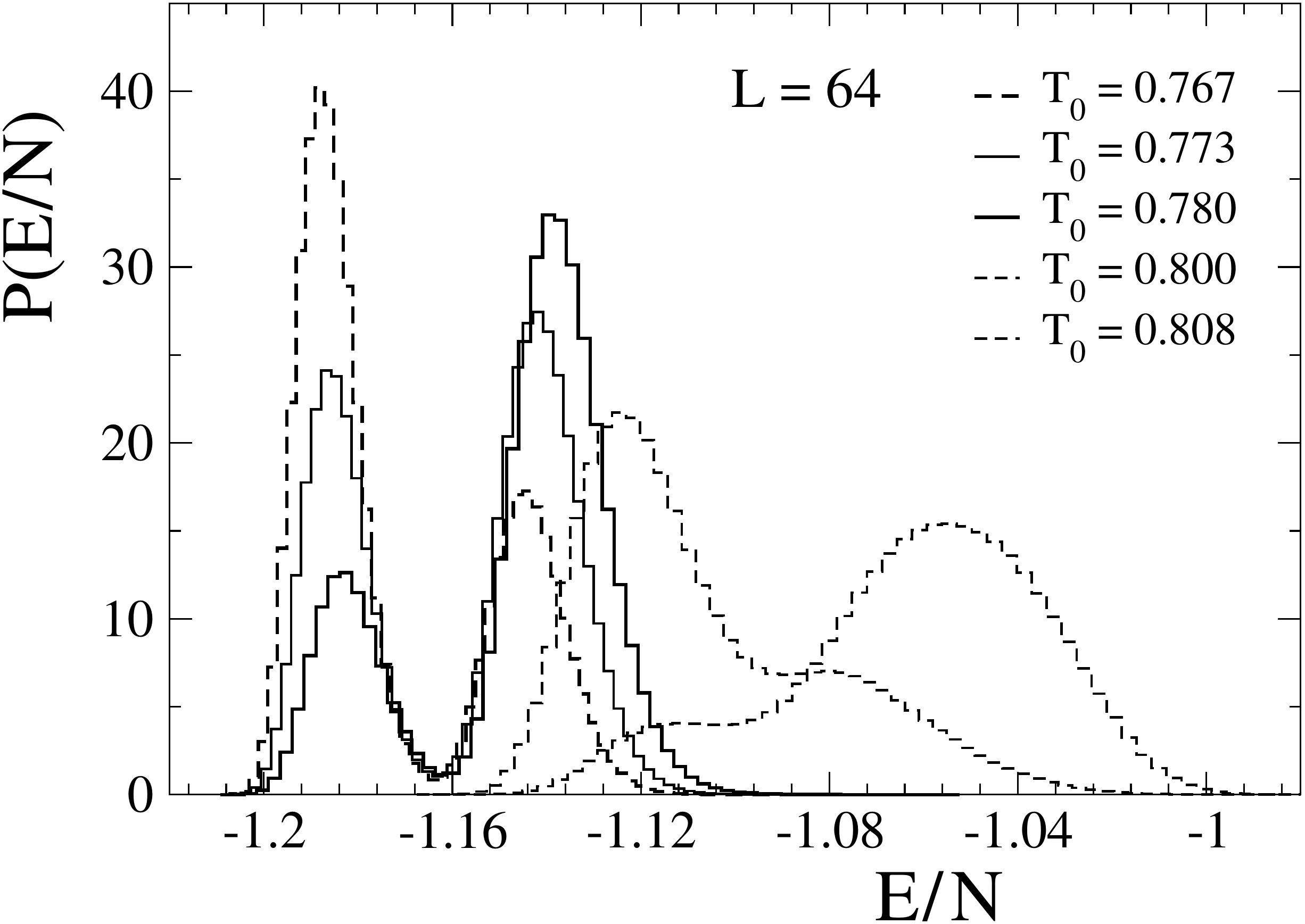}
\end{minipage}%
\caption{Energy density distributions for $L=64$ obtained from simulations at temperatures $T_0$.}
\label{fig:6}
\end{center}
\end{figure}

   Now, looking at the normalized energy density distributions for the second transition, Fig. 5(b) and (d),
we also observe the characteristic double-peak behavior although the histograms in Fig. 5(b) have less pronounced peaks. 
   We observe higher peaks for $L=64$ and 72 after reweighting to the critical point where
 $P_L(E_{nem}/N,T_h) = P_L(E_{tetra}/N,T_h)$.
   For this transition one obtains $f_0 = 0.0077(2)$  and $f_1= -0.269(23)$. 
   The second transition is clearly weaker compared with the first one. 
   Just to place some comparative figures to clarify the meaning of these numbers, one has $f_0 = 0.04735$ for 2D Potts model with 10 states and $f_0 = 0.010395$ for this model with 7 states. 
   These numbers are exact results.
   The first-order character of the model with 10 states is easily identified and is sufficient to work with
lattice sizes up to $L^2 \approx 50$.
   It is well known that this size is too small for such identification in the case of $q=7$.
   
   Long-range interactions always impose strong finite-size effects.
   Although our analysis procedure based on reweighting techniques with multiple histograms extracts the maximum information available in the MC data, some aspects still prevent us from definitive conclusions about the nature of the phase transitions.
   In all cases, the bulk correlation length $\xi$ plays an important role.
   Actually, $\xi$ and the extension of finite systems $L$ are the only relevant length scales.
   The validity of the FSS ansatz requires that $L > \xi$.
   The first transition seems to be of KT type or even of first-order nature.
   For a KT type transition, it is predicted that $\xi$ has the following behavior \cite{kost_74} at $T_c$
\begin{equation}
  \xi  \sim \exp( b/t^{1/2})  \, ,
\end{equation}
where $t= T/T_c -1$ is the reduced temperature.
   For first-order transitions  $\xi$ remains finite and for second-order ones it behaves as  $\xi  \sim t^{-\nu}$.
   Thus, at $T_c$ the correlation length diverges exponentially, while divergence is the power law in a second-order phase transition.
    Unfortunately, a dedicated work would be required to clarify such scenario \cite{gupta_92}.
    If present, the KT type transition would be responsible for eliminating the long-range order in the
striped phase through the formation of unbound dislocations \cite{abanov_54_1995}.
    The reason for considering a KT type transition is the absence of divergences or significant finite-size effects in the response functions.
    However, this absence may be justified by the enormous difficulty in sampling
configurations at this transition, as revealed later in our analysis of integrated autocorrelation times.

\begin{figure}[!t]
\begin{center}
\begin{minipage}[h]{0.70\textwidth}
\includegraphics[width=0.8\textwidth]{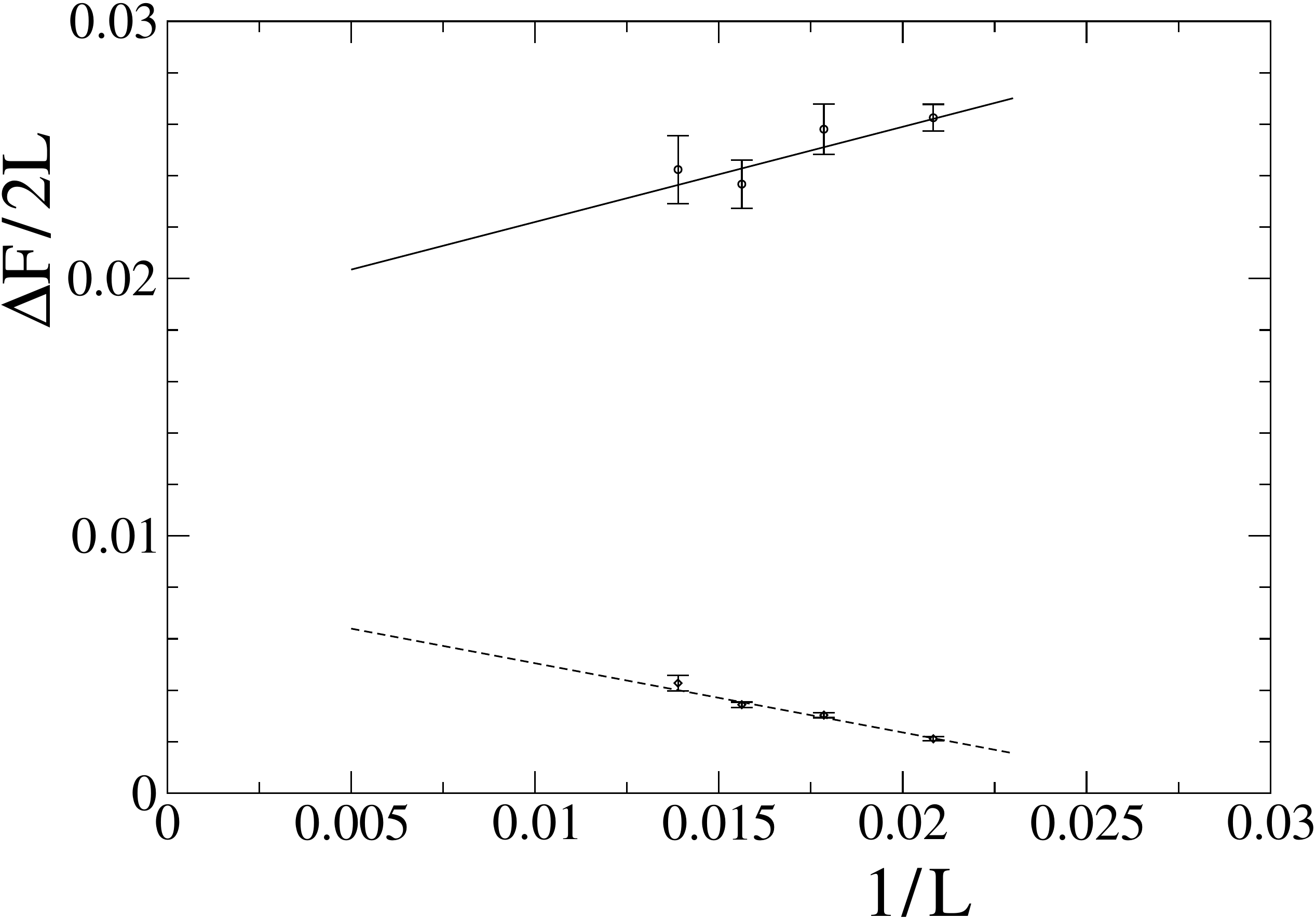}
\end{minipage}
\caption{Extrapolation of the free energy as a function of $L^{-1}$ for the striped-nematic  (---) and 
  nematic-tetragonal (- - -) transitions.}
\label{fig:7}
\end{center}
\end{figure}

   Our results from the interface tension analysis for the nematic-tetragonal transition indicate a very weak first-order phase transition.
   This transition has been described as second-order from the FSS analysis of the response functions specific heat and susceptibility. 
   Therefore, the first-order character exhibited by the histograms for the larger lattice sizes $L=64$ and 72
does not seem to be caught by this FSS analysis.  
   This implies that the fluctuations in the energy and order parameter are somehow damped in our data.
   This may result from the small lattice sizes we work with, where the true first-order character
is not effective and, presumably, combined with the need of a larger number of independent configurations in both phases.
   Actually, according to our results the effect of the free-energy barrier should be more severe between the striped and nematic phases.
   In turn, reliable calculation of the free-energy barriers requires enough configurations at $E_{m}$ that are exponentially suppressed by the Boltzmann factor in canonical MC simulations.

\section{Integrated autocorrelation time}

  Local MC updates are not efficient, and the produced data correspond to successive states of a Markov chain.   
  This chain may be highly correlated, introducing bias to the estimates of the physical quantities and unreliable variances if the sampling is not large enough.
  Also, related to systems with long-range interactions, the presence of a very slow approach to equilibrium introduces a further degree of complexity. 
  This has been highlighted in a recent study by Cannas {\it et al.} \cite{cannas_78_2008}
on the dipolar model.
  They found stronger metastabilities associated with the striped-nematic transition compared
with the nematic-tetragonal one.
  Here, we complement our study by evaluating the integrated autocorrelation time, to assert the effectiveness of the collected data at each temperature $T_0$ with local update algorithms.
  To this end, we define the normalized autocorrelation function for the energy time series \cite{alves_376_1990,sokal_89},
\begin{eqnarray}
 \rho(i) & = & \frac{1}{\sigma^2(E)} \, 
       \langle(E_s -\langle E \rangle)(E_{s+i} -\langle E \rangle)\rangle    \nonumber \\
         & = & \frac{1}{\sigma^2(E)_{(n_{\tau}-i)}}\,
               \sum_{s=1}^{n_{\tau}-i}(E_s-\langle E\rangle)(E_{s+i}-\langle E\rangle) \, ,
\end{eqnarray}
 where $\sigma(E)$ is the energy standard deviation. 
 The integrated autocorrelation time $\tau_{int}$,
\begin{equation}
 \tau_{int}  = \frac{1}{2} \rho(0) + \sum_{i=1}^{n_{\tau}} \rho(i)
\end{equation}
estimates the number of independent data points in a long sequence of $n_{\tau}$ MC measurements. 
 Actually, twice this value is the proper quantity to effectively evaluate this independence.
 It is usually expected that the autocorrelation function decays as a power law.
 However, this behavior is changed by a temperature dependent factor  near critical points
\cite{sokal_89}.

\begin{figure}[t]
\begin{center}
\begin{minipage}[h]{0.50\textwidth}
\subfigure{\includegraphics[width=0.95\textwidth]{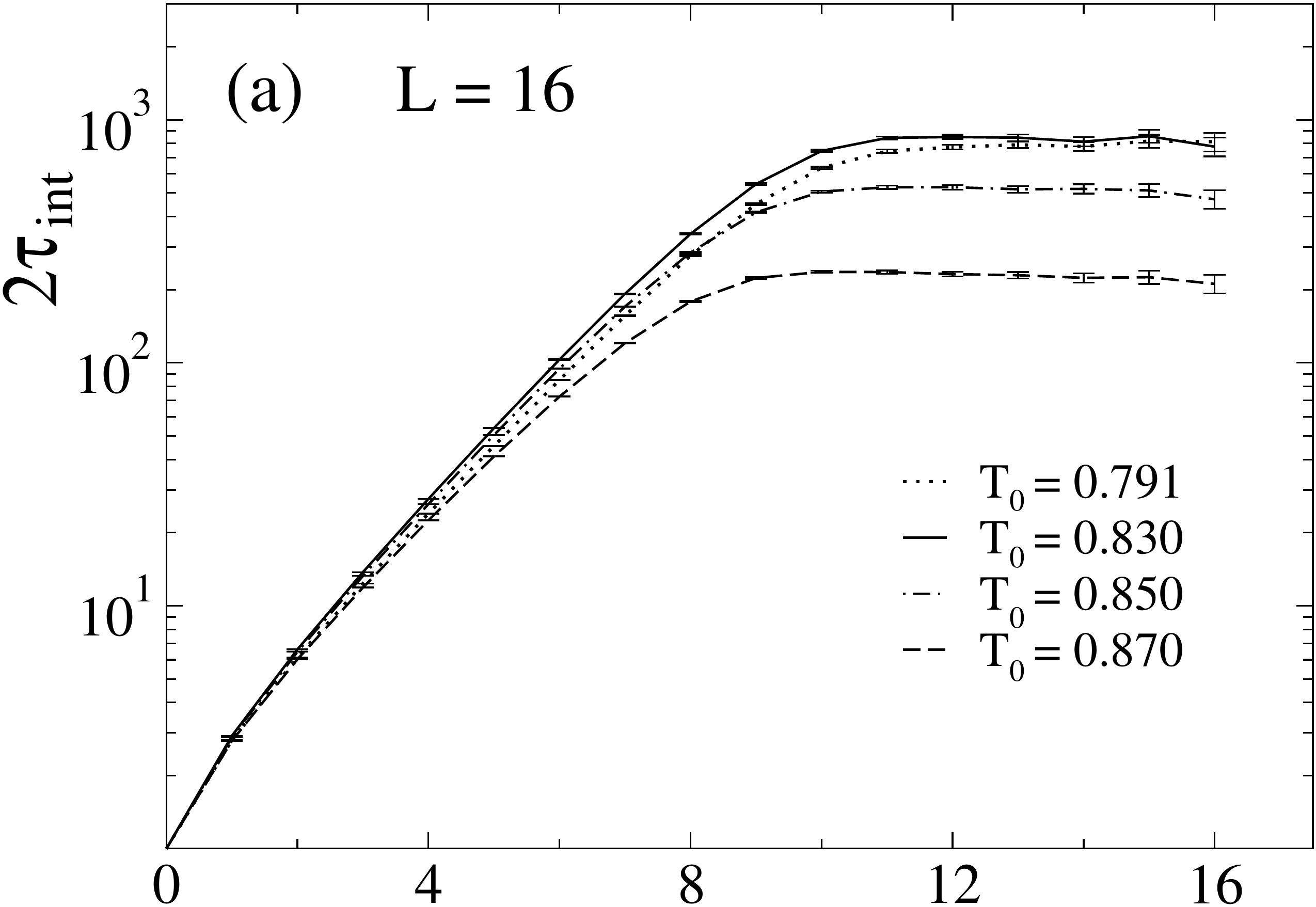}}
\end{minipage}%
\begin{minipage}[h]{0.50\textwidth}
\subfigure{\includegraphics[width=0.92\textwidth]{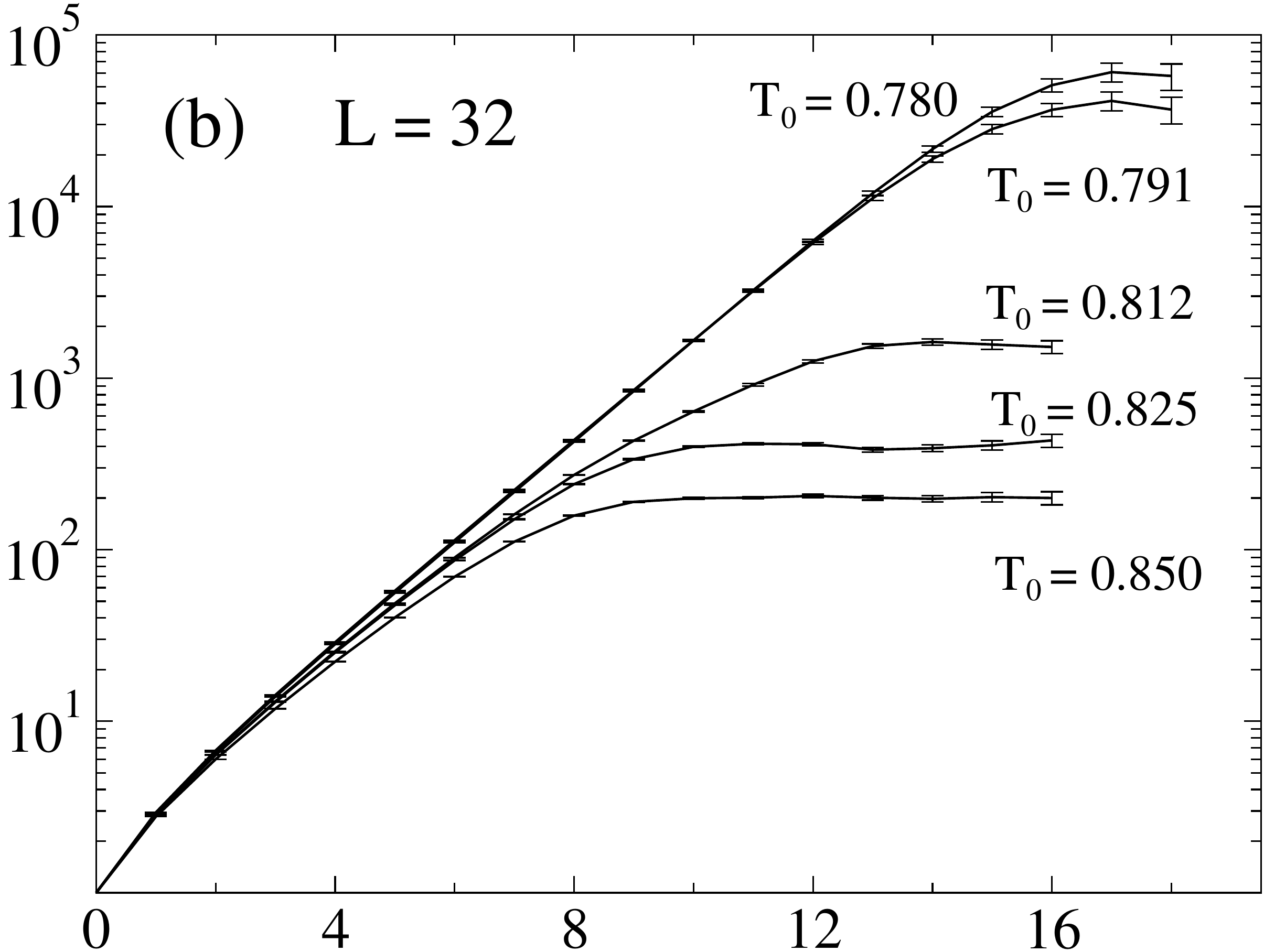}}
\end{minipage}

\vspace{0.4cm}

\begin{minipage}[h]{0.50\textwidth}
\subfigure{\includegraphics[width=0.95\textwidth]{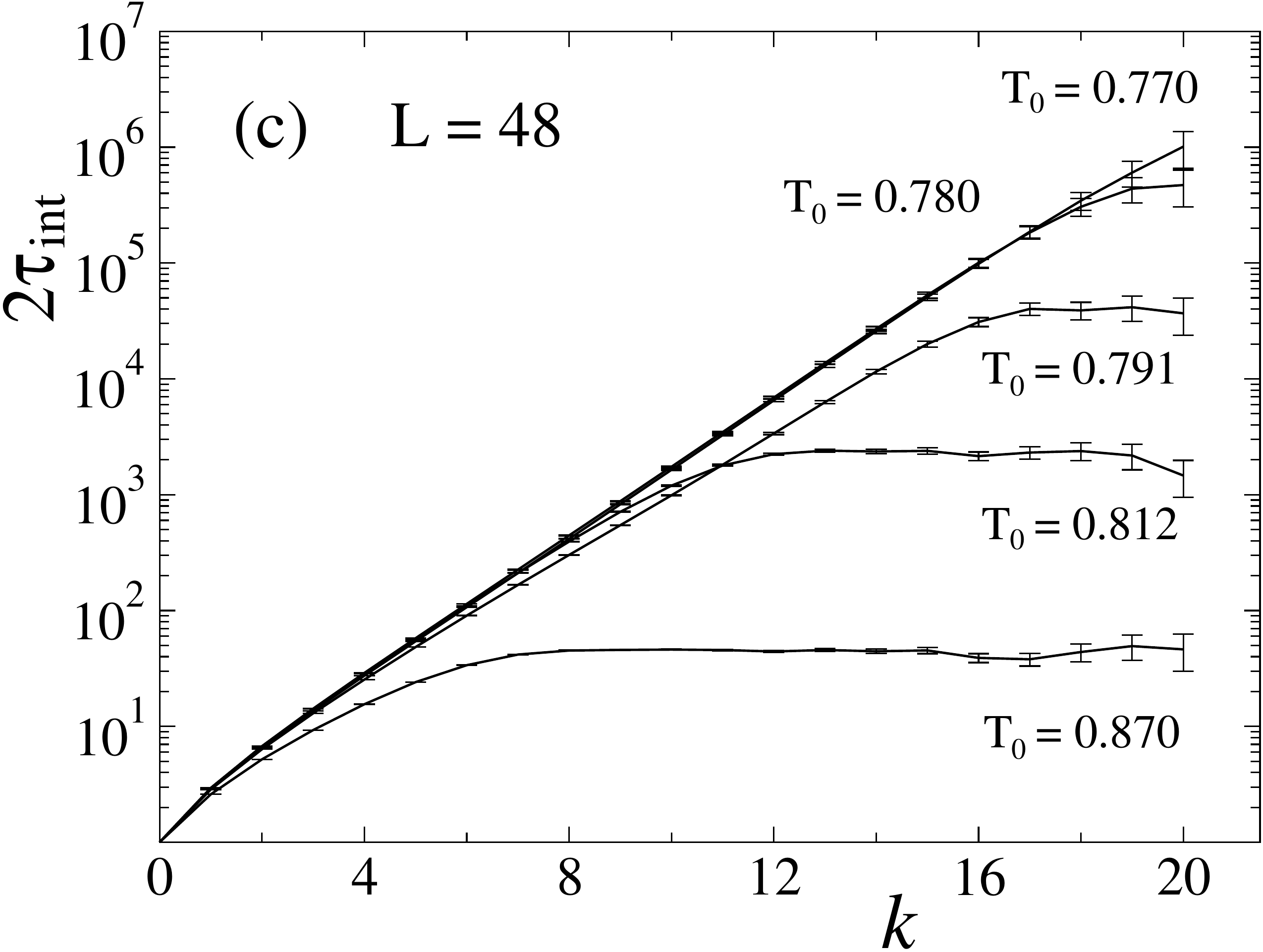}}
\end{minipage}%
\begin{minipage}[h]{0.50\textwidth}
\subfigure{\includegraphics[width=0.92\textwidth]{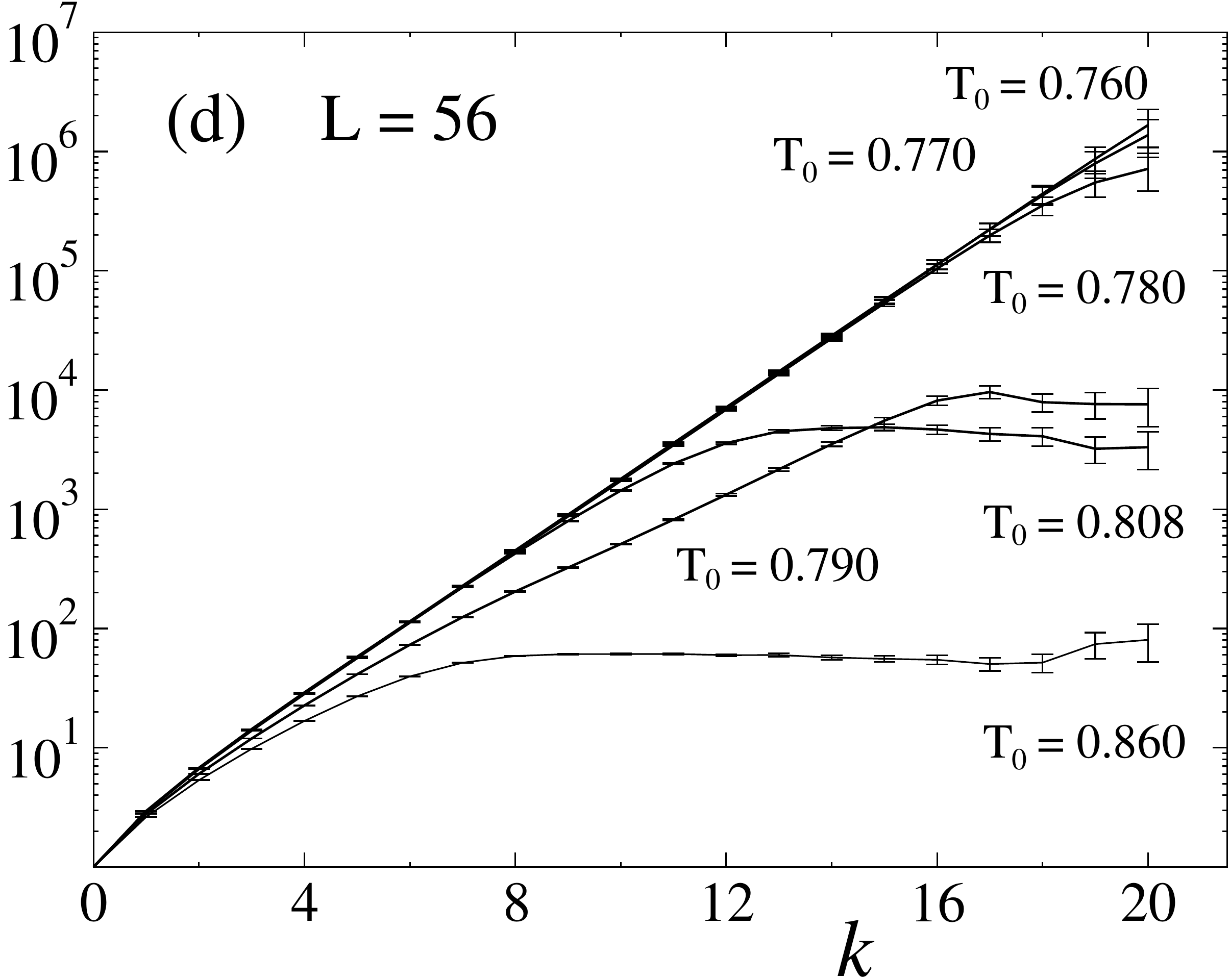}}
\end{minipage}
\end{center}
\caption{Integrated autocorrelation time $2\tau_{int}$ for lattice sizes $L=16$ (a), $L=32$ (b),
$L=48$ (c) and $L=56$ (d) as a function of time series length $2^k$.}
\label{fig:8}
\end{figure}

 Our calculations of $2 \tau_{int}$ are shown in Fig. 8 for lattice sizes $L= 16, 32, 48$ and $56$,
where we have introduced the notation $ k$ to describe the sample lengths, $n_{\tau}= 2^k$.
 
 We obtain $2\tau_{int} = 800 $ measurements ($L=16$) for temperatures close to the finite-size
critical point $T_c = 0.832$ and smaller values for temperatures in the tetragonal phase.
 For the $L=32$ case, Fig. 8(b) shows a larger value:  $2\tau_{int} = 6 \times 10^4$  
for $T= 0.780$ in the striped phase and $4 \times 10^4$ for  $T=0.791$, in the vicinity of the
specific heat maximum.
  Again, as expected, smaller values are observed for samples in the tetragonal phase.

  Figure 8(c) displays the results for $L=48$, whose specific heat data present two critical temperatures. 
  For the temperature inside the striped phase ($T= 0.770$), one obtains a larger 
integrated autocorrelation time compared with the one achieved at the striped-nematic transition
$2\tau_{int}(T=0.780) = 5 \times 10^5$.
  In the nematic phase, one obtains $2\tau_{int}(T=0.791) = 4 \times 10^4$.
  By increasing the temperature toward the next critical point $(T_c= 0.8132)$, 
one obtains $2\tau_{int}(T=0.812) =  2.4 \times 10^3$. 
  As the temperature increases further, it is observed a strong decrease in $2\tau_{int}$, reaching values as small as 45 for $T=0.870$.

  Figure 8(d) shows huge numbers for this autocorrelation time, larger than $2 \times 10^6$ for 
 data collected in the striped and at the striped-Ising nematic transition.
  In the Ising nematic phase, one has $2\tau_{int}(T=0.790) = 9.6 \times 10^3$.
  At the Ising nematic-tetragonal transition one obtains  $2\tau_{int}(T=0.808) = 5 \times 10^3$,
and a smaller value 60  at the temperature $T=0.860$ is achieved in the tetragonal phase.

   Here, we could try to describe autocorrelation times for the energy observable as a simple
power law  $ 2\tau_{int,E} = A L^{z_{int}}$, where $z_{int}$ is the  associated dynamic critical exponent.
   A rough estimate gives $z_{int} \approx 6.2$ and $z_{int} \approx 4.6$
for the striped-nematic and nematic-tetragonal phase transitions, respectively.
   These are large numbers since local MC algorithms generally have a dynamic critical exponent 
$z \approx 2$ \cite{sokal_B961}.
   For non-local update algorithms, such as the Swendsen-Wang cluster algorithm,
this exponent is further reduced to $z = 0.222(7)$ in the pure 2D Ising model.

  From the autocorrelation time analysis, we may infer the existence of very long-lived states near and at the striped-nematic transition.
  These states may lead to stronger bias in the determination of the response functions at this transition 
 than at the nematic-tetragonal transition, where $\tau_{int}$ are orders of magnitude smaller.

\section{Conclusions}

  Our systematic analysis by reweighting techniques in multiple histograms describes the existence of two phase transitions, which correspond to a two-step melting process leading to a disordered state.
  This long-range interaction model is described by a complex phase diagram with lines corresponding
to striped-nematic and nematic-tetragonal phase transitions, depending on the coupling $\delta$.
  We have seen that the true phenomenology of this model can be observed for large lattice sizes only, and that the controversial nematic phase is detected in a narrow range of temperatures.
  As $\delta$ increases, one observes larger stripe width.
  As realized for $h=2$, we also expect that for larger values of $h$, it will be necessary to increase further the system size to obtain reliable results.
  This can be explained by the need of inserting dislocations in the striped-domain structure.
  The formation of bound dislocation pairs proliferates as the temperature increases, leading to a phase transition.

  Our numerical results clearly show the locations of the phase transitions, which are strongly lattice size dependent. 
  Moreover, the response functions also have strong finite-size effects, which frustrate any convincing evaluation of the critical exponents from simple FSS analysis.
  
  The histogram analysis has shown a strong first-order character for the striped-nematic transition.
  However, this character may introduce strong metastabilities, leading to bias in our simulation.
  This is because the sampling may spend most of the time in one of two phases, mainly in the case of our largest lattice sizes.
  According to our integrated autocorrelation time calculations, we have learnt that to obtain $10^3$ independent configurations for $L=72$, about $10^{10}$ and $10^{7}$ MC sweeps are required for the striped-nematic and nematic-tetragonal transitions, respectively.
  We have attenuated this limited sampling by matching simulations at different temperatures but close to each other.
  However, it is not just a matter of how long one needs to perform the simulations.
  Our analysis has also revealed the need for larger lattice sizes, so that the character of the phase transitions is exposed.
  It is likely that the lattice sizes are not large enough compared to the correlation lengths.
  This because our critical exponent estimates from FSS indicate continuous transitions, in contrast with the results for the free-energy barriers.
  The asymptotic behaviors of the free-energy barriers between domains indicate that the interface tensions do not vanish in the thermodynamic limit. 

  Another controversial point still remains, the possibility of a KT type transition for the nematic-tetragonal transition.
  Our results for the response functions do not discard this possibility.
 
  The integrated autocorrelation time calculations show stronger critical slowing down in both
phase transitions when compared to the pure 2D Ising model.
  From this analysis we may conclude that there are very long-lived states near and at the
striped-nematic phase transition as opposed to the nematic-tetragonal transition, where
$\tau_{int}$ present smaller orders of magnitude.
  These results are very important to figure out how reliable the MC sampling for this system 
 with long-range interaction is. 
  Moreover, the use of important techniques for MC data analysis establishes the limits one can reach with local
update algorithms.
  More efficient results to obtain a deeper understanding of the complex striped-nematic-tetragonal phase transitions
seem possible with a variant of Wollf cluster algorithm \cite{luijten} or by means of generalized ensemble methods.

\section{Acknowledgments}
The authors acknowledge support by FAPESP and CAPES (Brazil).


\end{document}